\documentclass[twocolumn]{aastex63}
\usepackage{amsmath}
\usepackage{comment}

\newcommand{\be}{\begin{equation}}
\newcommand{\ee}{\end{equation}}
\newcommand{\pfrac}[2]{\left( \frac{#1}{#2} \right)}

\shorttitle{QPEs from Star-Disk Collisions in Galactic Nuclei}
\shortauthors{Itai Linial, B.D.~Metzger}
\graphicspath{{./}{figures/}}

\begin{document}

\title{EMRI + TDE = QPE:\\
Periodic X-ray Flares from Star-Disk Collisions in Galactic Nuclei}

\author[0000-0002-8304-1988]{Itai Linial}
\affil{Institute for Advanced Study, 1 Einstein Drive, Princeton, NJ 08540, USA}

\author[0000-0002-4670-7509]{Brian D.~Metzger}
\affil{Department of Physics and Columbia Astrophysics Laboratory, Columbia University, New York, NY 10027, USA}
\affil{Center for Computational Astrophysics, Flatiron Institute, 162 5th Ave, New York, NY 10010, USA} 

\correspondingauthor{Itai Linial}
\email{itailin@ias.edu}

\begin{abstract}
Roughly half of the quasi-periodic eruption (QPE) sources in galactic nuclei exhibit a remarkably regular alternating ``long-short'' pattern of recurrence times between consecutive flares. We show that a main-sequence star (brought into the nucleus as an extreme mass-ratio inspiral; EMRI) which passes twice per orbit through the accretion disk of the supermassive black-hole (SMBH) on a mildly eccentric inclined orbit, each time shocking and ejecting optically-thick gas clouds above and below the midplane, naturally reproduces observed properties of QPE flares. Inefficient photon production in the ejecta renders the QPE emission much harder than the blackbody temperature, enabling the flares to stick out from the softer quiescent disk spectrum. Destruction of the star via mass ablation limits the QPE lifetime to decades, precluding a long-lived AGN as the gaseous disk. By contrast, a tidal disruption event (TDE) naturally provides a transient gaseous disk on the requisite radial scale, with a rate exceeding the EMRI inward migration rate, suggesting that many TDEs should host a QPE. This picture is consistent with the X-ray TDE observed several years prior to the QPE appearance from GSN 069. Remarkably, a second TDE-like flare was observed from this event, starting immediately after detectable QPE activity ceased; this event could plausibly result from the (partial or complete) destruction of the QPE-generating star triggered by runaway mass-loss, though other explanations cannot be excluded. Our model can also be applied to black hole-disk collisions, such as those invoked in the context of the candidate SMBH binary OJ 287.

\end{abstract}


\section{Introduction}

Quasi-periodic eruptions (QPEs) are short ($\lesssim$ hours-long) X-ray transients spatially coincident with galactic nuclei that recur on a timescale of several hours to almost a day and which exhibit peak luminosities $\gtrsim 10^{42}$ erg s$^{-1}$ in the 0.5-2 keV X-ray band at least an order of magnitude higher than the quiescent X-ray luminosity \citep{Miniutti+19,Giustini+20,Arcodia+21,Chakraborty+21,Arcodia+22,Miniutti+23,Webbe&Young23}.  Their spectra appear quasi-thermal with temperatures in the range $\approx 100-200$ eV, corresponding to blackbody emission radii comparable to a solar radius (e.g., \citealt{Krolik&Linial22}; however, see \citealt{Miniutti+23}).

The stellar masses of QPE host galaxies are low $\lesssim 10^{10}M_{\odot}$ \citep{Miniutti+19,Arcodia+21}, suggesting that their nuclei likely host supermassive black holes (SMBH) with commensurably low masses $M_{\bullet} \sim 10^{5}-5\times 10^{6}M_{\odot}$ \citep{Miniutti+19,Giustini+20,Wevers+22}.  The hosts exhibit no canonical active galactic nuclei (AGN)-like broad emission lines nor any infrared photometric excess indicating the presence of hot dust \citep{Miniutti+19,Arcodia+21}.  However, the optical spectra of the first two QPE sources, GSN 069 and RXJ1301.9+2747, show narrow line emission lines with clear AGN-driven ionization \citep{Miniutti+13,Sun+13}; \citet{Wevers+22} also found evidence for narrow-line AGN in the hosts of the two QPEs discovered in the eROSITA survey \citep{Arcodia+21}.  Nevertheless, the lack of a broad-lined region or substantial optical/UV continuum disfavors the presence of radially-extended AGN disks.  

Several theoretical models have been proposed to explain the QPE phenomenon, including (i) accretion disc instabilities  \citep{Sniegowska+22,Raj&Nixon21,Pan+22,Kaur+22}; (ii) gravitational lensing of a SMBH binary \citep{Ingram+21}; (iii) mass transfer onto a SMBH from one or more orbiting bodies \citep{Zalamea+10,King20,Zhao+22,King22,King23,Chen+22,Metzger+22,Wang+22,Krolik&Linial22,Linial&Sari22,Lu&Quataert22}; and (iv) collisions between an orbiting secondary body and the SMBH accretion disk \citep{Sukova+21,Xian+21,Franchini+23}.  We focus on the latter scenario here.

An important clue to the nature of QPEs is a remarkably regular alternating pattern in some sources, in which the time interval separating consecutive bursts varies back and forth between two values which differ by about 10\%. In addition, flares that precede longer recurrence intervals appear to be systematically brighter than those appearing before short ones \citep[e.g.,][]{Miniutti+23}. This oscillating ``long/short'' behavior is observed only in GSN 069 \citep{Miniutti+19} and eRO-QPE2 \citep{Arcodia+21},
leading \citet{Arcodia+22} to suggest the existence of two sub-classes of QPE behavior: those which exhibit the regular long-short oscillation behavior and those showing less regular burst timing (i.e., the remaining confirmed systems eRO-QPE1 and RXJ 1301.9+2747; \citealt{Giustini+20,Arcodia+21,Arcodia+22}).  Some properties of the regular, ``long/short alternating'' QPE sources are summarized in Table \ref{tab:sources}.

From a purely phenomenological perspective, the long-short recurrence behavior can be naturally explained in a scenario whereby QPEs are generated by the collisions between a mildly eccentric inclined orbiting body and a gaseous accretion disk \citep{Sukova+21,Xian+21,Franchini+23}, as occur twice per orbit.\footnote{\citet{King23} proposes that rapid mass transfer from a white dwarf onto the SMBH can introduce oscillatory behavior about the mean; however, it is not clear to us how variable mass-transfer could produce the observed large $\sim 10\%$ back-and-forth difference in the flare arrival times.}  Although several studies have been dedicated to the hydrodynamics and radiation of the collision between a black hole and the disk of a more massive SMBH (e.g., \citealt{Lehto&Valtonen96,Subr&Karas99,Semerak+99,Pihajoki+16}), in part motivated by the SMBH binary candidate OJ 287 (e.g., \citealt{Sillanpaa+88,Valtonen+08,Komossa+23}), or to star-disk collisions (e.g., \citealt{Zentsova83,Vokrouhlicky&Karas93,Nayakshin+04,Dai+10,MacLeod&Lin20,Sukova+21}), only a few of these works make concrete predictions for the flare emission properties. 

\citet{Nayakshin+04} calculate the relatively weak X-ray flares generated by a star passing through a hypothesized low-mass gas disk surrounding Sgr A$^{*}$; however, some of the assumptions behind their calculations, such as the neglect of radiation trapping and adiabatic losses, are not valid for collisions in the inner regions of relatively high$-\dot{M}$ disks relevant to QPE sources.  \citet{Sukova+21} perform 2D and 3D MHD simulations of star-disk interactions, focusing on the influence of the orbiting body on the ejection of plasma from the disk and the time-variability of its accretion rate, finding this interaction to be a promising mechanism for generating QPE-like time-structure.  However, they do not make an explicit calculation of the radiation from this interaction; their calculations also focus on thick, radiatively inefficient disks whose emission properties differ from the quiescent emission seen from at least some QPE sources (e.g., \citealt{Miniutti+23}).  \citet{Lehto&Valtonen96} predict the flare emission from the collision of a secondary SMBH with the accretion disk of the primary SMBH in the context of OJ 287; however, they make several assumptions that differ substantially from what we find in this work.  

A related clue to the origin of QPEs comes from the long-term behavior of the quiescent X-ray flux of the well-studied source GSN 069 (see \citealt{Miniutti+23} for a review).  While QPEs were observed from GSN 069 for approximately a year starting in Dec.~2018 (but could have started as early as 2015; \citealt{Miniutti+19}), the same system exhibited starting earlier in 2010 a several year X-ray outburst  \citep{Saxton+11,Miniutti+13} consistent in luminosity $L_{\rm X} \sim 7\times 10^{43}$ erg s$^{-1}$ with being a tidal disruption event (TDE) \citep{Shu+18,Sheng+21}, though exhibiting a somewhat longer duration and slower post-maximum decay than most TDE flares.\footnote{ Since producing two QPE-like outbursts in 2006, the candidate QPE-source XMMSL1 J024916.6-041244 \citep{Chakraborty+21} has exhibited a long-term $L_{\rm X} \propto t^{-5/3}$ decay in its quiescent X-ray light curve, compatible with a TDE being the source of the gaseous disk in this system as well.}  Moreover, after QPE activity ceased to be detected near the end of 2019, the same system exhibited a {\it second} high-amplitude long-lived X-ray outburst with a qualitatively similar light curve shape,
consistent with being another stellar-disruption in the same galactic nucleus \citep{Miniutti+23}.  While the fallback debris from the first TDE provides a natural explanation for the gaseous disk required for QPE emission via star-disk collisions \citep{Miniutti+19}, a second TDE would on face be highly surprising because the average interval between TDEs in a typical galactic nucleus is $\gtrsim 10^{4}$ yr (e.g., \citealt{Stone&Metzger16,Yao+23}).  If the first TDE was a partial disruption (e.g., \citealt{Nixon&Coughlin22,Bortolas+23}), then the second outburst could conceivably be powered by the disruption of the surviving core \citep{Miniutti+23}.   Here, we will offer an alternative explanation for the second high-amplitude accretion-powered event, which has the added benefit of naturally predicting a $\sim$decade-long delay following the first TDE and contributing to its atypically shallow decay. 
 
 Furthermore, while it is tempting to attribute the QPE-generating body orbiting the SMBH to be the surviving stellar core of a partial TDE \citep{King20,Sheng+21,Xian+21,Zhao+22}, the core will generally inherit the highly eccentric orbit of the original star if not become even more eccentric due to the kick received during the disruption process (e.g., \citealt{Manukian+13,Gafton+15,Cufari+23}; see discussion in \citealt{Metzger+22}), inconsistent with the mildly eccentric orbit needed to explain the long-short QPE behavior \citep{Xian+21}.  As in previous related models \citep{Metzger+22,Linial&Sari22,Krolik&Linial22,Lu&Quataert22}, we find it more natural to consider the QPE-generating star as one that has separately migrated to the galactic center in an extreme mass-ratio inspiral (EMRI; e.g., \citealt{Linial&Sari17,Linial&Sari22}), independent of the disk-generating TDE.  

In this paper we show that the periodic collisions between a stellar EMRI and an SMBH accretion disk naturally generate flares with properties consistent with observed QPEs.  We start in Sec.~\ref{sec:stardisk} by considering the general interaction between a star and a gaseous disk of a fixed accretion rate.  After concluding that a QPE-generating star would not survive long interacting with a sustained AGN disk, we specialize in Sec.~\ref{sec:TDE} to a transient gaseous disk created by a tidal disruption event (TDE).  In Sec.~\ref{sec:applications} we consider applications of our model, particularly to the well-studied QPE source GSN 069; our model can also be applied to black hole-disk collisions, so we also consider implications for the periodically flaring SMBH binary candidate OJ 287.  In Sec.~\ref{sec:conclusions} we summarize our conclusions.

\begin{table*}
\begin{center}
    \caption{Properties of Candidate Long/Short Oscillating QPE Sources}
    \begin{tabular}{|l|c|c|c|c|c|c|c|c|c|}
    \hline
     Source & $P_{\rm QPE}^{(a)}$ & $L_{\rm QPE}^{(b)}$ & $t_{\rm QPE}^{(c)}$ & $k_{\rm B}T_{\rm obs}^{(d)}$ & $L_{\rm Q}^{(e)}$ & $T_{\rm Q}^{(f)}$ & $M_{\bullet}$ &  $\mathcal{D}^{(g)}$ & Reference \\
     \hline
- & (hr) & (erg s$^{-1}$) & (hr) & (eV) & (erg s$^{-1}$) & (eV) & ($M_{\odot}$) & - & - \\
      \hline \hline
GSN-069 & $8.5$-$9.5$ & $1.3\times 10^{42}$ & 1.1 & $100$-$120$ & $\sim 10^{43}$  & 50 & $10^{5.5}$-$10^{6.5}$ & 0.12 & 1 \\
eRO-QPE2 & $2.36$-$2.75$ & $1.5\times 10^{42}$ & 0.18 & 190-240 & $\gtrsim 10^{41}$ & 75 & $10^{5}$-$10^{6}$  & 0.07 & 2 \\
\hline 
    \end{tabular}\\
    $^{(a)}$QPE period, with the given range indicating the amplitude of the ``long-short'' recurrence time behavior; $^{(b)}$ Characteristic X-ray luminosity of QPE flare, roughly equal to the fluence divided by the characteristic duration $\sim t_{\rm QPE}$.  We estimate this as $e^{-1}$ of the peak bolometric luminosity, quiescent emission subtracted; $^{(c)}$Duration of QPE flare, estimated as twice the FWHM; $^{(d)}$Spectral temperature of QPE flare, quiescent emission subtracted; $^{(e)}$Quiescent X-ray luminosity; $^{(f)}$Quiescent X-ray temperature; $^{(g)}$Flare duty cycle $\equiv t_{\rm QPE}/P_{\rm QPE}$.
    $^{1}$\citet{Miniutti+19,Miniutti+23,Wevers+22}
    $^{2}$\citet{Arcodia+21,Arcodia+22}
    \label{tab:sources}
    \end{center}
\end{table*}

\section{Disk-Star Interaction}
\label{sec:stardisk}

We first review the basic properties of the SMBH accretion disk, which generates the quiescent X-ray emission in our model (Sec.~\ref{sec:disk}), before moving onto the properties of the stellar EMRI and its orbit (Sec.~\ref{sec:star}).  In Sec.~\ref{sec:collisions} we discuss the star-disk collisions and their expected electromagnetic emission, before addressing the physical processes responsible for limiting the QPE lifetime in Sec.~\ref{sec:lifetime}.  

\subsection{The Quiescent Disk}
\label{sec:disk}

We consider an SMBH of mass $M_{\bullet} = 10^{6}M_{\bullet,6}M_{\odot}$ accreting gas steadily at a rate $\dot{M} = \dot{m}\dot{M}_{\rm Edd}$, where
$\dot{M}_{\rm Edd} \equiv L_{\rm edd}/(\epsilon c^{2}) \approx 1.7\times 10^{24}\,{\rm g\,s^{-1}}M_{\bullet,6}$ is the Eddington accretion rate for a characteristic radiative efficiency of $\epsilon = 0.1$.  For typical values $10^{-3} \lesssim \dot{m} \lesssim 1$ corresponding to radiatively efficient accretion and the observed quiescent luminosities of QPE sources, radiation pressure dominates over gas pressure in the disk midplane and the vertical aspect ratio at radii $r \gg R_{\rm g}$ can be written (e.g., \citealt{Frank+02}) 
\be
\frac{h}{r} \simeq \frac{3}{2\epsilon}\frac{R_{\rm g}}{r}\frac{\dot{M}}{\dot{M}_{\rm Edd}} \simeq 1.5\times 10^{-2} \; \dot{m}_{-1}\left(\frac{r}{100R_{\rm g}}\right)^{-1},
\label{eq:hoverr}
\ee
where $h$ is the vertical scale-height, $R_{\rm g} \equiv GM_{\bullet}/c^{2}$ and $\dot{m} = 0.1\dot{m}_{-1}$.  The stability of radiation dominated accretion disks remains a long-standing question (e.g., \citealt{Lightman&Eardley74,Hirose+09,Jiang+13,Jiang+19}), and their vertical structure likely deviates significantly from that predicted by an $\alpha$-disk model (e.g., \citealt{Blaes+06}); however, insofar as the QPE emission properties in our model are not particularly sensitive to the vertical extent of the disk, we neglect these complications.

The optical depth through the disk midplane of surface density $\Sigma \simeq \dot{M}/(3\pi \nu)$ can be written
\be
\tau_{\rm c} = \Sigma \kappa_{\rm T} \simeq \frac{\dot{M}\kappa_{\rm T}}{3\pi \nu} \approx \frac{6.0\times 10^{3}}{\alpha_{-1}\dot{m}_{-1}}\left(\frac{r}{100R_{\rm g}}\right)^{3/2},
\label{eq:tauc}
\ee
where $\kappa_{\rm T} \simeq 0.34$ cm$^{2}$ g$^{-1}$ is the electron scattering opacity, $\nu = \alpha (GM_{\bullet} r)^{1/2}(h/r)^{2}$ is the kinematic viscosity \citep{Shakura&Sunyaev73}, and we scale $\alpha = 0.1\alpha_{-1}$ to a characteristic value.  The midplane temperature is given by
\begin{eqnarray}
k_{\rm B} T_{\rm c} &=& k_{\rm B} \left(\frac{3\tau_{\rm c}}{2a\kappa_{\rm T}}\frac{ GM_{\bullet}}{r^{2}}\frac{h}{r}\right)^{1/4} \nonumber \\
&\approx& 37 \,{\rm eV}\, \,\alpha_{-1}^{-1/4}M_{\bullet,6}^{-1/4}\left(\frac{r}{100R_{\rm g}}\right)^{-3/8},
\label{eq:Tc}
\end{eqnarray}
where $k_{\rm B}$ is the Boltzmann constant and $a$ is the radiation constant.  Absent interaction with orbiting bodies (i.e., when in ``quiescence''), the disk emission is dominated by radii near the innermost circular orbit $R_{\rm isco}$, with total luminosity
\be
L_{\rm Q} = \dot{m}L_{\rm Edd} \simeq 1.5\times 10^{43}\,{\rm erg\,s^{-1}}\dot{m}_{-1}M_{\bullet,6}
\label{eq:LQ},
\ee
and characteristic emission temperature
\be
k_{\rm B}T_{\rm Q} \approx k_{\rm B}\left(\frac{3GM_{\bullet}\dot{M}}{8\pi \sigma R_{\rm isco}^{3}}\right)^{1/4} \simeq 59\,{\rm eV}\,\frac{\dot{m}_{-1}^{1/4}}{M_{\bullet,6}^{1/4}}\left(\frac{R_{\rm isco}}{4R_{\rm g}}\right)^{-3/4},
\label{eq:TQ}
\ee
consistent with the quiescent emission temperatures of QPE sources (Table \ref{tab:sources}).

\subsection{The Stellar EMRI}
\label{sec:star}

We consider a star of radius $R_{\star} = \mathcal{R}_{\star} R_{\odot}$ and mass $M_{\star} = \mathcal{M}_{\star} M_{\odot}$ on a nearly-circular orbit around the SMBH of semi-major axis $r_{0}$ and orbital velocity
\be
v_{\rm K} \approx \left(\frac{GM_{\bullet}}{r_{0}}\right)^{1/2} \approx 0.10\,{\rm c}\,\left(\frac{r_{0}}{100R_{\rm g}}\right)^{-1/2}.
\label{eq:vK}
\ee
If the QPE period, $P_{\rm QPE}$, is set by collisions between the star and the disk which occur twice per orbital period $P_{\rm orb} \simeq 2\pi(r_0^{3}/GM_{\bullet})^{1/2}$, then 
\be
P_{\rm QPE} = \frac{P_{\rm orb}}{2} \simeq 4.3\,{\rm hr}\,M_{\bullet,6}\left(\frac{r_{0}}{100R_{\rm g}}\right)^{3/2}.
\label{eq:TQPE}
\ee
Insofar as we are considering mildly eccentric orbits $e \lesssim 0.1$ appropriate to observed long-short alternating QPE, the value of $P_{\rm QPE}$ here  can roughly be taken to be the average interval between consecutive flares.  Explaining observed QPE periods $P_{\rm QPE} \approx 2-19$ hr (Table \ref{tab:sources}) thus requires $r_{0}/R_{\rm g} \approx (60-300)M_{\bullet,6}^{-2/3}$. 
In what follows we shall express key results in terms of the directly observable $P_{\rm QPE} = 4$ hr $\mathcal{P}_{\rm QPE,4}$.

As we shall discuss in Sec.~\ref{sec:why}, the star likely reached this location through gradual gravitational wave inspiral as a stellar-EMRI \citep{Linial&Sari17,Linial&Sari22}.  Thus, at the time of the observed QPE emission, the star may either be overflowing its Roche lobe ($r_{0} \approx r_{\rm T}$) undergoing mass transfer onto the SMBH \citep{Linial&Sari17}, or may yet have to enter Roche contact ($r_{0} > r_{\rm T}$), where
\begin{eqnarray}
r_{\rm T} &\simeq& R_{\star}\left(\frac{M_{\bullet}}{M_{\star}}\right)^{1/3} \approx 7\times 10^{12}\,{\rm cm}\,\mathcal{R}_{\star}\mathcal{M}_{\star}^{-1/3}M_{\bullet,6}^{1/3} 
\label{eq:Rt}
\end{eqnarray}
is the tidal radius.  The condition $r_0 \ge r_{\rm T}$ defines a minimum QPE period in this scenario
\be
P_{\rm QPE,min} \simeq \pi \left(\frac{R_{\star}^{3}}{GM_{\star}}\right)^{1/2} \simeq 1.4\,{\rm hr}\,\mathcal{R}_{\star}^{3/2}\mathcal{M}_{\star}^{-1/3}\,.
\label{eq:TQPEmin}
\ee
In Sec.~\ref{sec:why} we show that QPE flares are detectable over the disk quiescent emission only for orbits with $r_0$ moderately larger than $r_{\rm T}$ ($P_{\rm QPE} \gtrsim P_{\rm QPE,min}$).

\begin{figure*}
    \centering
    \includegraphics[width= \textwidth]{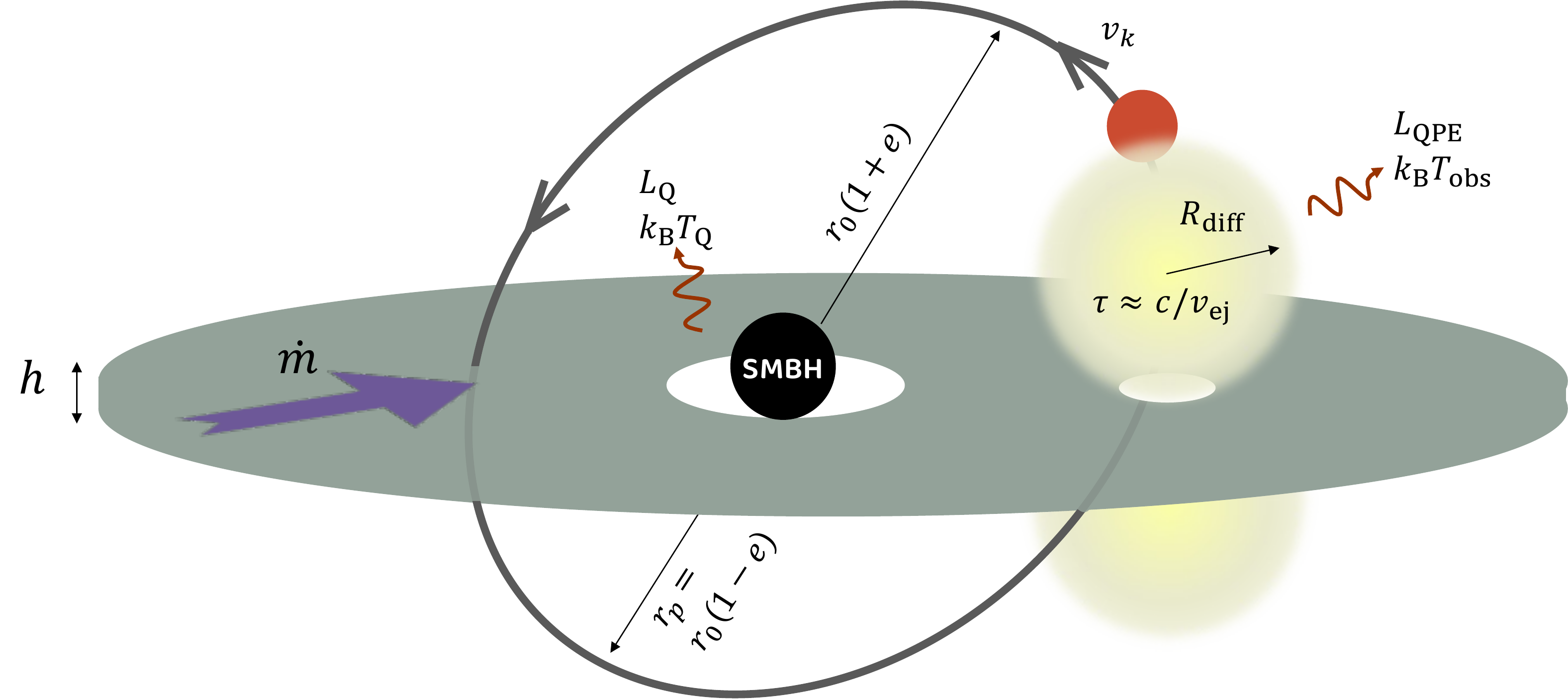}
    \caption{Schematic view of our model. A star orbits an SMBH that is accreting matter through a thin disk of scale height $h$ at a rate $\dot{m}$. Due to the inclined orbital plane, the star impacts the disk twice per orbit, carving a hole through the disk and ejecting an optically thick cloud of material expanding above and below the disk plane.  As the ejecta expands and cools, photons begin to diffuse out and the light curve peaks once the optical depth drops below $c/v_{\rm ej}$, where $v_{\rm ej} \sim v_{\rm K}$ is the ejecta velocity imparted by the colliding star. The inner regions of the disk dominate the soft quiescent emission seen between the collision-powered flares.}
    \label{fig:cartoon}
\end{figure*}

 The star's orbital plane must be significantly misaligned with that of the accretion disk to generate QPE emission, with the observed alternating long-short recurrence time pattern explained in part by the star spending a longer time on the side of the disk near apocenter than on the pericenter side (\citealt{Miniutti+19,Xian+21}; see Fig.~\ref{fig:cartoon} for a schematic illustration).
 
In addition to the very gradual orbital decay due to gravitational wave emission (Sec.~\ref{sec:rates}) and gas drag (Sec.~\ref{sec:dragdecay}), the star's orbit is subject to more rapid evolution as a result of other general relativistic effects (e.g., \citealt{Xian+21,Metzger+22,Franchini+23}).  The fastest of these is apsidal precession, which can lead to secular evolution of the long-short recurrence time difference amplitude.  Given the characteristic precession angle per orbit $
\delta \epsilon \simeq 6\pi(R_{\rm g}/r_{0})$, significant precession  ($\Delta \epsilon \sim 2\pi$) will occur on a timescale,
\be
T_{\epsilon} \sim \frac{\Delta \epsilon}{\delta \epsilon}P_{\rm orb} \simeq \frac{P_{\rm orb}}{3}\frac{r_{0}}{R_{\rm g}} \approx 10.6 \; {\rm d} \, M_{\bullet,6}^{-2/3} \mathcal{P}_{\rm QPE,4}^{5/3},
\label{eq:Tepsilon}
\ee
sufficiently short to be observed in QPE light curve epochs spanning months \citep{Xian+21,Franchini+23}.  

Nodal precession can also occur, potentially leading to changes in the inclination angle between the orbital plane and the accretion disk.  At leading post-Newtonian order, nodal precession is driven by Lense-Thirring frame dragging, with significant nodal precession ($\Delta \Omega \sim 2\pi$) thus occurring on a timescale
\be
T_{\Omega} \sim \frac{\Delta \Omega}{\delta \Omega}P_{\rm orb} \simeq \frac{P_{\rm orb}}{2a_{\bullet}}\left(\frac{r_{0}}{R_{\rm g}}\right)^{3/2} \approx 155\,{\rm d} \,\,\frac{\mathcal{P}_{\rm QPE,4}^{2}}{a_{\bullet}M_{\bullet,6}}
\label{eq:TLT},
\ee
where $\delta \Omega = 4\pi a_{\bullet}(r_{0}/R_{\rm g})^{-3/2}$ is the per-orbit nodal shift and $0 \le a_{\bullet} \le 1$ is the dimensionless spin magnitude of the SMBH \citep{Merritt10}.  If the SMBH spin axis is misaligned with angular momentum axis of the stellar orbit, the orbit can come in and out of alignment with the disk midplane on a timescale as short as a year.

In addition to variations in the timing of the flares due to the evolving geometry of the orbit with respect to the disk plane, light travel times from the two collision sites introduces an additional source of timing variations. The magnitude of this effect is roughly of order
\be
    \frac{2r_0/c}{P_{\rm QPE}} \approx 6\times 10^{-2} \; M_{\bullet,6}^{-2/3}\mathcal{P}_{\rm QPE,4}^{-1/3}  \,,
\ee
not much smaller than the observed variations in the flare timing.

\subsection{Star-Disk Collisions}
\label{sec:collisions}

\begin{deluxetable}{cc}
\tablecaption{QPE Flare Model Parameters}
\tablewidth{700pt}
\tabletypesize{\scriptsize}

\label{tab:params}
\tablehead{
\colhead{Symbol} & \colhead{Description} 
} 
\startdata
$M_{\bullet} = M_{\bullet,6} 10^{6}M_{\odot} $ & SMBH mass \\
$\dot{m} = 0.1\dot{m}_{-1}$ & SMBH Eddington accretion rate \\
$\alpha = 0.1\alpha_{-1}$ & Disk viscosity parameter \\
$P_{\rm QPE} = $ 4 hr $\mathcal{P}_{\rm QPE,4}$ & QPE period ($=P_{\rm orb}/2$; Eq.~\eqref{eq:TQPE}) \\
$R_{\star} = \mathcal{R}_{\star}R_{\odot}$ & Stellar EMRI radius  \\
$M_{\star} = \mathcal{M}_{\star}M_{\odot}$ & Stellar EMRI mass  \\
\enddata
\end{deluxetable}

\subsubsection{Emission from Shocked Disk Ejecta}
\label{sec:lightcurve}

The star will twice per orbit pass through the disk midplane of thickness $h/R_{\star} \simeq 3.2 \, \dot{m}_{-1}M_{\bullet,6}/\mathcal{R}_{\star}$ similar to the stellar radius.  For simplicity we consider a nearly head-on collision (i.e., a 90 degree angle between the angular momenta of the orbit and disk) in what follows.  Assuming that the disk has returned to an unperturbed state by the time of each collision (a condition we shall check in Sec.~\ref{sec:hole}), the mass of the disk material intercepted each passage,
\begin{eqnarray}
M_{\rm ej} \simeq 2\pi R_{\star}^{2}\Sigma \approx 2.4\times 10^{-7} \; {\rm M_{\odot}} \, \frac{\mathcal{R}_{\star}^{2} \mathcal{P}_{\rm QPE,4}}{\alpha_{-1}\dot{m}_{-1} M_{\bullet,6}},
\label{eq:Mej}
\end{eqnarray}
is much less than the star's mass, where we have used Eq.~\eqref{eq:tauc} for the disk surface density.  We have augmented the swept-up mass by a factor of 2 relative to the star's cross section $\pi R_{\star}^{2}$ to account for the transverse motion of the disk, which moves into the star at the same speed as the star moves through the disk.

The mass $M_{\rm ej}$ will be shocked by the star, and the resulting high pressure will cause the matter to re-expand both above and below the disk midplane\footnote{Similar to the process illustrated by Harold Edgerton's famous images of a bullet piercing an apple.} (e.g., \citealt{Lehto&Valtonen96,Ivanov+98,Pihajoki+16,Sukova+21}), acquiring an outwards velocity $v_{\rm ej} \approx v_{\rm sh}$ as well as thermal and kinetic energy in roughly equal parts,
\begin{align}
    E_{\rm ej} \simeq \frac{1}{2}M_{\rm ej} v_{\rm sh}^{2} \approx
    4.5\times 10^{45}\,{\rm erg}\,\frac{\mathcal{R}_{\star}^{2}\mathcal{P}_{\rm QPE,4}^{1/3}}{\alpha_{-1}\dot{m}_{-1}M_{\bullet,6}^{1/3}},
    \label{eq:Eej}
\end{align}
where $v_{\rm sh} \simeq \sqrt{2}v_{\rm K}$ is the shock speed, taken to be the relative velocity between the star and disk.

Supported by numerical simulations of black hole-disk (e.g., \citealt{Ivanov+98}) and star-disk (e.g., \citealt{Sukova+21}) collisions, we assume the ejecta which emerges from each side of the disk spreads spherically in all directions from the break-out point (a region on the surface of the disk of area $\simeq \pi R_{\star}^{2}$), with the ejecta radius measured from this point $R_{\rm ej} = v_{\rm ej}t$ at times $t > 0$ from the breakout.  Such ballistic expansion at close to the sound speed of the freshly-shocked gas is achieved because (1) the shocked expanding layer is highly opaque, such that its thermal energy will mostly be converted into kinetic energy via PdV work over just a few expansion times, before a significant fraction of the remaining energy escapes as radiation (as is also the case in supernovae); (2) the ejecta radius at the time of peak light is typically $\ll r_0$ (see below), such that the influence of the gravitational acceleration of the SMBH on the debris evolution can be neglected to first order.  

At times $t \gg R_\star/v_{\rm ej}$, the optical depth of the spherically expanding ejecta declines as
\be
\tau(t) \approx \frac{\kappa M_{\rm ej}}{4\pi R_{\rm ej}^2} \propto t^{-2} \,,
\label{eq:tauej}
\ee
where $\kappa \approx \kappa_{\rm T}$ is the ejecta's opacity.  Similar to a supernova, photons will escape once $\tau(t_{\rm QPE}) \approx c/v_{\rm ej}$, reached at time
\begin{multline}
t_{\rm QPE} \approx \left(\frac{\kappa M_{\rm ej}}{4\pi c v_{\rm ej}}\right)^{1/2} \approx 
0.09\,{\rm hr}\frac{\mathcal{R}_{\star}\mathcal{P}_{\rm QPE,4}^{2/3}}{\alpha_{-1}^{1/2}\dot{m}_{-1}^{1/2}M_{\bullet,6}^{2/3}},  \,
\label{eq:tpk}
\end{multline}
and corresponding radius $R_{\rm diff} \simeq v_{\rm K}t_{\rm QPE}.$ Within a factor of a few, Eq.~\eqref{eq:tpk} sets the rise time and decay time, as well as the overall duration, of the bolometric light curve (e.g., \citealt{Arnett80}).

Before it can be radiated, the thermal energy $E_{\rm ej}$ is reduced by adiabatic expansion, by a factor $(V_0 /4\pi R_{\rm diff}^3 )^{\gamma-1}$, where $V_0 \approx \pi R_\star^2 h/7$ is the initial volume of the material shocked by the star (where the factor of 7 accounts for the density compression ratio of the shocked gas), and $\gamma$ is the ejecta's adiabatic index. Since radiation pressure dominates over gas pressure, we take $\gamma = 4/3$ implying a characteristic luminosity
\begin{eqnarray}
L_{\rm QPE} &\approx& \frac{E_{\rm ej} (R_\star^2 h)^{1/3} }{3v_{\rm ej} P_{\rm QPE}^2} \approx \frac{L_{\rm Edd}}{3}\frac{(R_\star^2 h)^{1/3} }{r_{0}} \nonumber \\
&\approx& 3.5 \times 10^{41}\,{\rm erg\,s^{-1}}\,\mathcal{R}_{\star}^{2/3}M_{\bullet,6}\dot{m}_{-1}^{1/3}\mathcal{P}_{\rm QPE,4}^{-2/3},
\label{eq:Lpk}
\end{eqnarray}
where we have used Eqs.~\eqref{eq:vK}, \eqref{eq:tpk}, and in the second equality, Eq.~\eqref{eq:hoverr}.  Moderate differences in $L_{\rm QPE}$ are likely for different viewing angles or due to asymmetry in the mass ejection along the star's ingress versus egress direction through the disk \citep{Ivanov+98,Miniutti+23}, for instance due to the difference in the relative speed between the star and orbiting disk material.  We stress that $L_{\rm QPE}$ is only weakly dependent on the specific properties of the disk (through the $h^{1/3}$ dependence in the above expression).\footnote{Predictions for the disk scale-height under different assumptions (e.g., regarding the degree of magnetic pressure support) typically vary by less than a factor of a few (e.g., \citealt{Kaur+22}, their Fig.~1).} This is due to the two-fold role of the swept up mass, $M_{\rm ej}$ - first, in setting the initial energy budget of the expanding ejecta, $E_{\rm ej} \propto M_{\rm ej}$, and secondly, in setting its optical depth, $\tau \propto M_{\rm ej}$. In deriving $L_{\rm QPE}$, these two dependencies on $M_{\rm ej}$ cancel out.

This radiation is emitted over a timescale $\sim t_{\rm QPE}$, corresponding to a flare duty cycle
\be
\mathcal{D} = \frac{t_{\rm QPE}}{P_{\rm QPE}} \approx 0.022 \, \frac{\mathcal{R}_{\star}}{\alpha_{-1}^{1/2}\dot{m}_{-1}^{1/2} M_{\bullet,6}^{2/3}\mathcal{P}_{\rm QPE,4}^{1/3}}
 \,.
\label{eq:duty}
\ee

At peak light when $\tau \simeq c/v_{\rm ej}$, the energy density of the radiation within the ejecta shell is given by $u_{\gamma} \simeq L_{\rm QPE}\tau/(4\pi R_{\rm diff}^{2}c)$.  The blackbody temperature of the radiation is thus given by
\be
k_{\rm B} T_{\rm BB} \simeq k_{\rm B}\left(\frac{u_{\gamma}}{a}\right)^{1/4}
\approx 12.6\,{\rm eV} \frac{\alpha_{-1}^{1/4}\dot{m}_{-1}^{1/3}M_{\bullet,6}^{1/3}}{\mathcal{R}_{\star}^{1/3}\mathcal{P}_{\rm QPE,4}^{1/4}}   \,.
\label{eq:TBB}
\ee
This is too soft to explain the observed temperatures $T_{\rm eff} \approx 100-200$ eV of QPE flares (Table \ref{tab:sources}).

However, in general the temperature of the escaping radiation can be harder than $T_{\rm BB}$, if photon production is not sufficient to achieve thermal equilibrium in the ejecta shell on the timescale of the emission (e.g., \citealt{Weaver76,Katz+10,Nakar&Sari10}).  In particular, if $n_{\rm BB} \approx a T_{\rm BB}^{4}/3k_{\rm B}T_{\rm BB}$ is the number density of photons in thermal equilibrium and $\dot{n}_{\rm \gamma, ff}(T,\rho)$ is the photon production rate via bremsstrahlung emission\footnote{Free-free processes will dominate over bound-free because of the high ionization parameter $\xi \equiv L_{\rm QPE}/n_{\rm ej}R_{\rm diff}^{2} \sim 10^{4}$ erg cm s$^{-1}$  (see, e.g., \citealt{Nayakshin+04}, their Fig.~4), where $n_{\rm ej}\simeq \rho_{\rm ej}/m_p$ and we have estimated the ejecta density $\rho_{\rm ej} \simeq M_{\rm ej}/(4\pi R_{\rm diff}^{2})$; the photon to baryon density ratio $\xi \propto \rho^{1/3}$ is even greater at the higher densities $\gg \rho_{\rm ej}$ which dominate the photon creation.} at temperature $T$ and density $\rho$, then thermal equilibrium of the shocked gas will only be achieved for $\eta \ll 1$ where (\citealt{Nakar&Sari10}; their Eq.~9)
\begin{multline} \label{eq:eta}
\eta \equiv \frac{n_{\rm BB}}{t_{\rm cross}\dot{n}_{\rm \gamma, ff}(T_{\rm BB,sh},\rho_{\rm sh})} \\
\approx \frac{220\,\rm s}{t_{\rm cross}}\left(\frac{\rho_{\rm sh}}{10^{-10}\rm \,g\,cm^{-3}}\right)^{-2}\left(\frac{T_{\rm BB,sh}}{10\,\rm eV}\right)^{7/2} \\
\approx
2.46 \alpha_{-1}^{9/8}\dot{m}_{-1}^{5/4}M_{\bullet,6}^{13/6}\mathcal{P}_{\rm QPE,4}^{-49/24}.
\end{multline}
Here, $t_{\rm cross} = (h/7)/v_{\rm K}$ is the time the freshly shocked gas spends in the immediate post-shock region of thickness $\approx h/7$, and in the final line we have evaluated the photon production rate using the density $\rho_{\rm sh} \simeq 7\rho_{\rm c}$ and blackbody temperature $T_{\rm BB,sh} \approx (3\rho_{\rm c}v_{\rm K}^{2}/a)^{1/4}$ of the shocked gas, where $\rho_{\rm c} \simeq \Sigma/(2h)$ is the unshocked midplane density and the factor of $7$ accounts for the shock compression.  We have used the immediate post-shock values here because the total photon production ($\dot{n}_{\rm \gamma, ff} \propto \rho^{2}T^{-7/2} \propto \rho^{5/6}$) in the adiabatically-expanding radiation-dominated ejecta ($\rho \propto T^{1/3}$) is dominated by small radii $\ll R_{\rm diff}$ where $\rho$ is greatest. We have also checked that the photons created in the freshly shocked material dominate those already present in the radiation-dominated disk; this is because the shocks heat matter to a specific internal energy $\sim v_{\rm K}^{2} \gg c_{\rm s}^{2}$ much greater than that of the unperturbed midplane, where $c_{\rm s} \simeq (h/r)v_{\rm K}$ is the sound speed of the unperturbed disk midplane of thickness $h/r \ll 1$.

For $\eta \gtrsim 1$, bremsstrahlung emission does not produce enough photons to achieve thermal equilibrium and the spectral temperature of radiation in the midplane $T_{\rm obs,c} \simeq \eta^{2}T_{\rm BB,c}$ exceeds the blackbody value \citep{Nakar&Sari10}.  The quadratic dependence $T_{\rm obs} \simeq \eta^{2}T_{\rm BB}$ follows because, for a fixed photon energy density $u_{\gamma} \approx n_{\gamma, \rm BB}(3kT_{\rm BB})$, the radiation temperature actually achieved $T_{\rm obs} \propto u_{\gamma}/n_{\gamma,\rm obs}$ scales inversely with the amount of free-free photon production $n_{\gamma, \rm obs} \propto \dot{n}_{\rm \gamma, ff}(T_{\rm obs}) \propto T_{\rm obs}^{-1/2}$, where $n_{\gamma, \rm BB}$ and $n_{\gamma, \rm obs}$ are the photon number densities at $T_{\rm BB}$ and $T_{\rm obs}$, respectively.  However, $\eta \propto 1/\dot{n}_{\rm \gamma, ff}(T_{\rm BB})$ is defined (Eq.~\eqref{eq:eta}) using the (higher) photon production rate at the blackbody temperature $T_{\rm BB} < T_{\rm obs}$, i.e. $\dot{n}_{\rm \gamma, ff}(T_{\rm obs}) = \dot{n}_{\rm \gamma, ff}(T_{\rm BB})(T_{\rm obs}/T_{\rm BB})^{-1/2}$. 

Insofar that adiabatic losses which occur between the midplane and the diffusion radius $R_{\rm diff}$ will act to reduce $T_{\rm obs}$ from $T_{\rm obs,c}$ by the same factor that $T_{\rm BB,c}$ is reduced to $T_{\rm BB}$, we infer that the observed spectral temperature will be given (for $\eta > 1$) by: 
\begin{eqnarray}    
    k_{\rm B}T_{\rm obs} &\approx&
    \eta^{2} k_{\rm B}T_{\rm BB} \nonumber \\
    &\approx&  {56 \, \rm eV} \, \frac{\alpha_{-1}^{5/2}\dot{m}_{-1}^{11/4}M_{\bullet,6}^{1/3}}{\mathcal{R}_{\star}^{1/3}}\left(\frac{r_0}{100R_{\rm g}}\right)^{-13/2} \nonumber \\
    &\approx& {76 \, \rm eV} \, \frac{\alpha_{-1}^{5/2}\dot{m}_{-1}^{11/4}M_{\bullet,6}^{14/3}}{\mathcal{R}_{\star}^{1/3} \mathcal{P}_{\rm QPE,4}^{13/3}} \,,
\label{eq:Tobs}
\end{eqnarray}
where $T_{\rm BB}$ is the blackbody temperature at the diffusion surface (Eq.~\eqref{eq:TBB}).

\begin{figure*}
    \centering
    \includegraphics[width=0.45\textwidth]{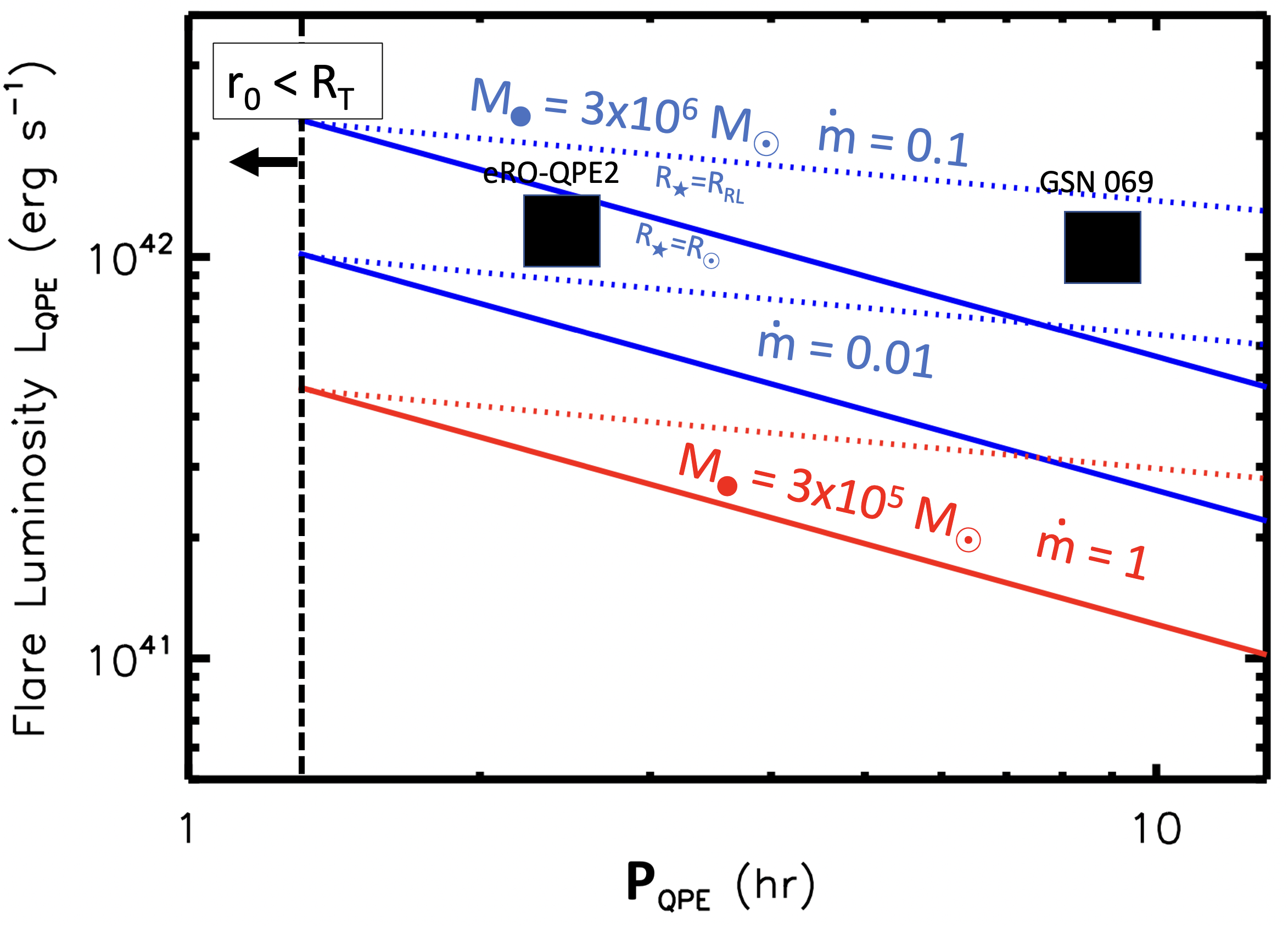}
    \includegraphics[width=0.45\textwidth]{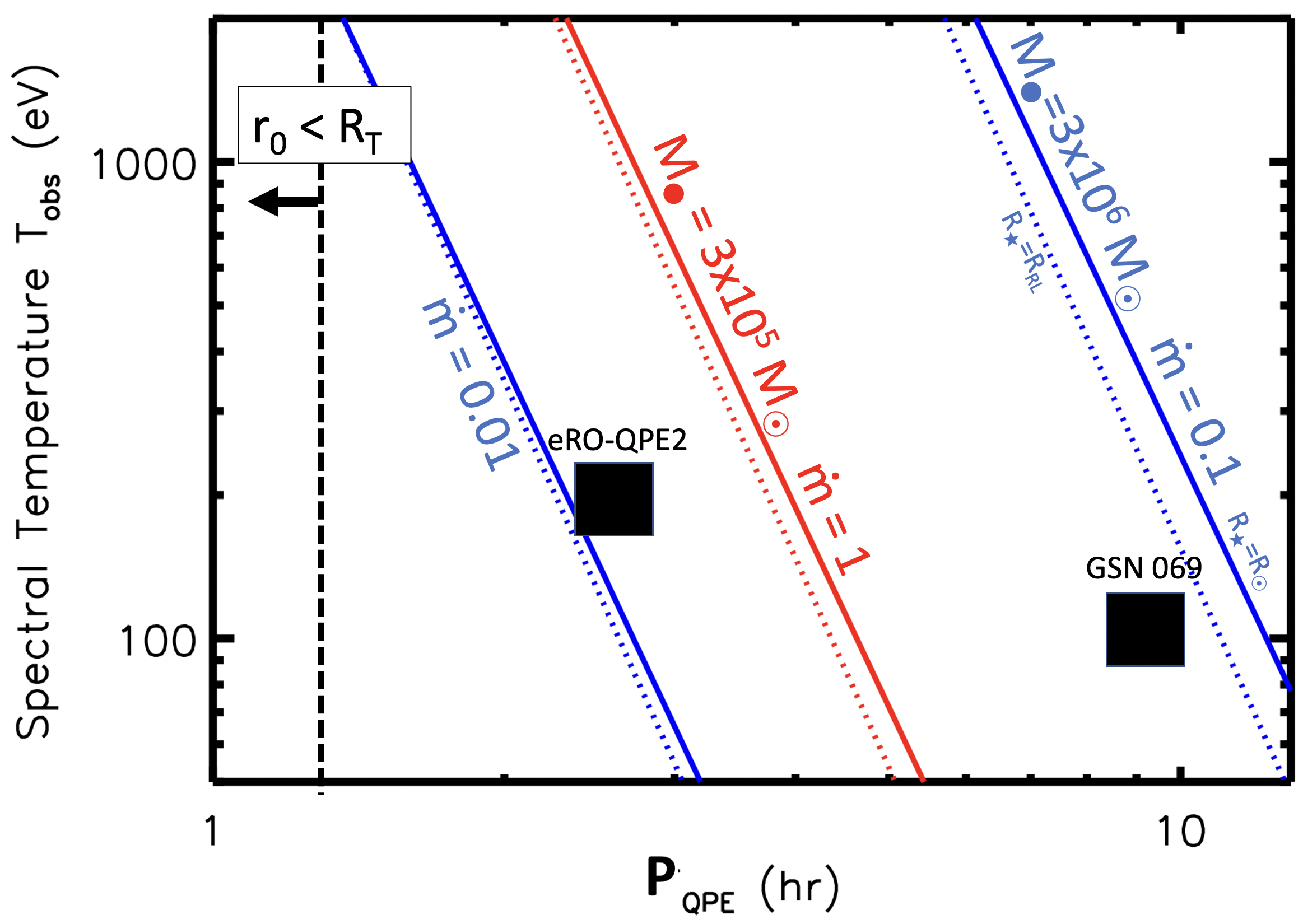}
    \includegraphics[width=0.45\textwidth]{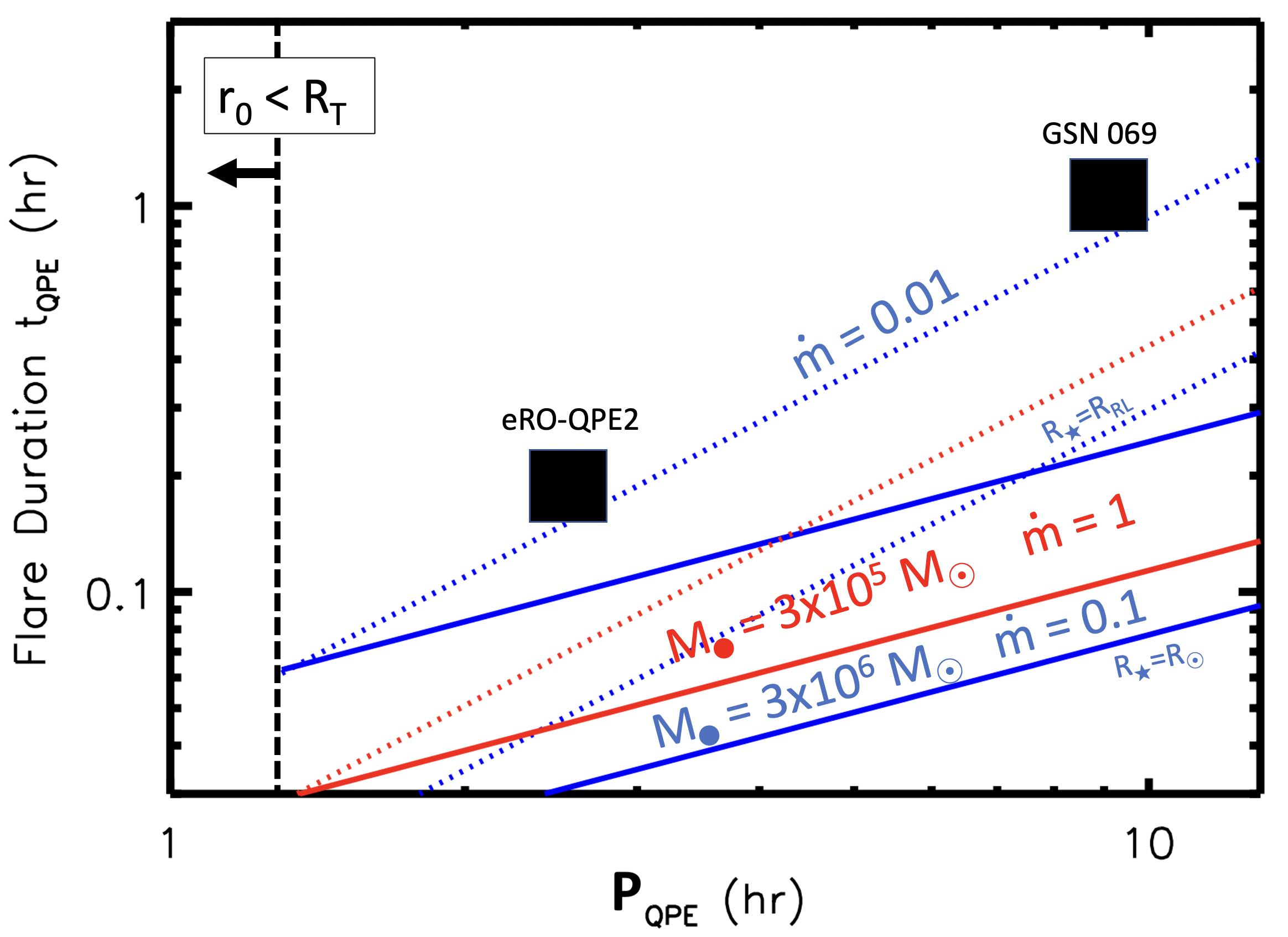}
    \includegraphics[width=0.44\textwidth]{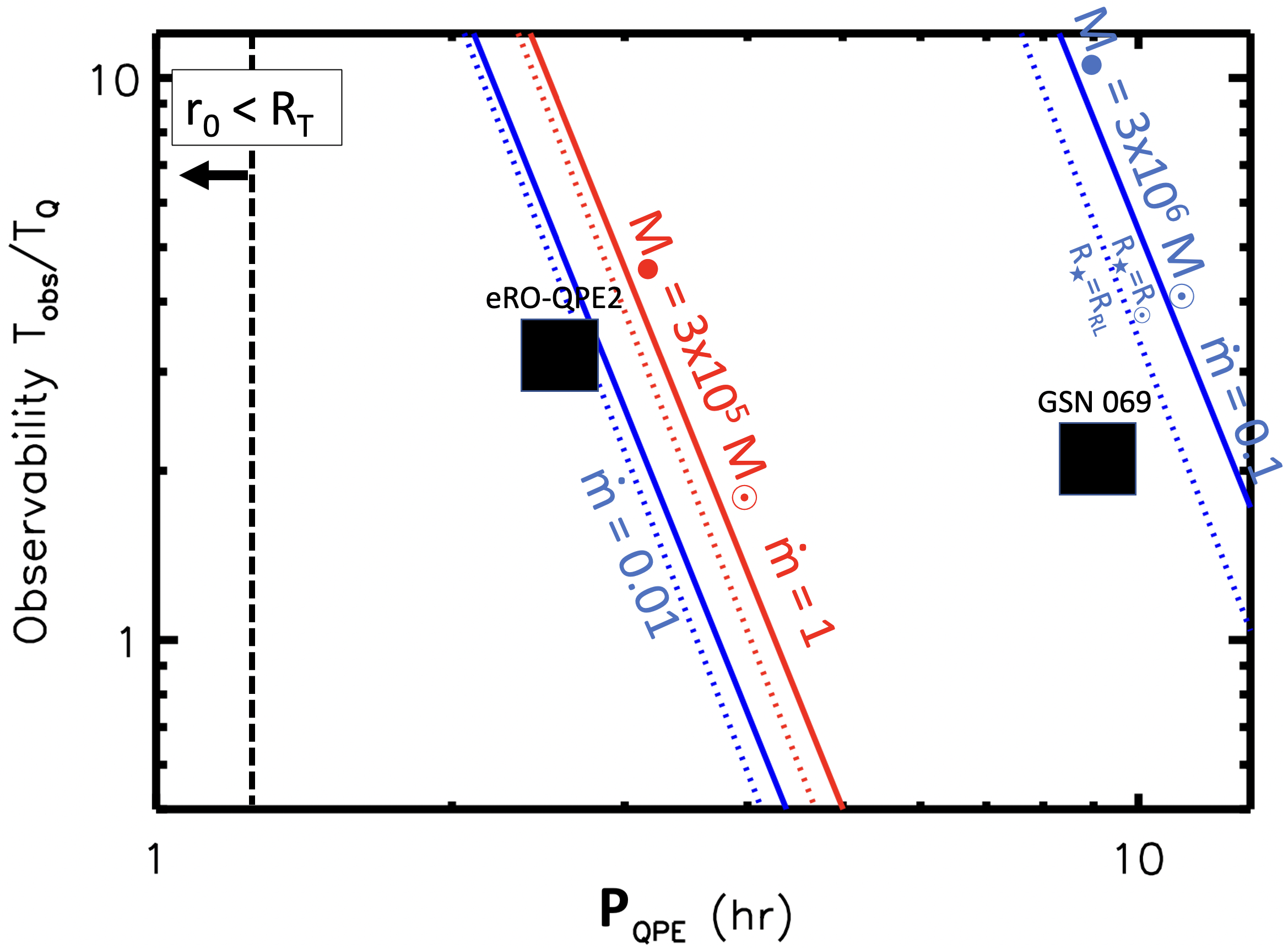}
    \caption{Characteristic luminosity (top left; Eq.~\eqref{eq:Lpk}), duration (bottom left; Eq.~\eqref{eq:tpk}), and spectral temperature (upper right; Eq.~\eqref{eq:Tobs}) of flares from disk-star collisions for a solar-mass EMRI for different SMBH masses and accretion rates as indicated.  The bottom right panel shows the ratio of the spectral temperature to that of the quiescent disk, $T_{\rm Q}$ (Eq.~\eqref{eq:TQ}), which may be taken as a metric of QPE observability.  Solid and dotted lines show separately cases where the stellar radius is taken to be a solar radius or equal to Roche lobe radius for overflow onto the SMBH, respectively; a vertical dashed line corresponds to the minimum QPE period for a star with $R_{\star} = R_{\odot}$ (Eq.~\eqref{eq:TQPEmin}).  Shown for comparison with black squares are the observed flare properties of long/short oscillating QPE (Table \ref{tab:sources}); the thickness assigned to the boxes are somewhat arbitrary and not to meant to encompass observational uncertainties or time-variation.  We assume $\alpha = 0.1$ and $R_{\rm isco} = 4 R_{\rm g}$ for the viscosity and inner radius, respectively, of the quiescent disk.}
    \label{fig:LpkTobs}
\end{figure*}

For characteristic parameters, we find that values $T_{\rm obs} \gtrsim 100$ eV, consistent with QPE observations can be achieved (see Fig.~\ref{fig:LpkTobs}). The predicted spectral shape from the $\tau \approx c/v_{\rm ej}$ surface is therefore that of thermal bremsstrahlung emission $F_{\nu} \propto \nu^{0}\exp\left[-h\nu/k_{\rm B}T_{\rm obs}\right]$, possibly modified by Comptonization.  Following Eqs.~12, 13 of \citet{Nakar&Sari10}, we estimate that Comptonization effects are likely to be mild for parameters characteristic of the freshly shocked gas.  However, if Comptonization effects do become important at early times, then the spectrum will be modified into a Wien spectrum $F_{\nu} \propto \nu^{3}\exp[-h\nu/k_{\rm B}T_{\rm obs}]$ at the highest frequencies (e.g., \citealt{Illarionov&Siuniaev75,Nakar&Sari10}).  We emphasize that although the emission surface is large $\sim R_{\rm diff} \sim 10^{12}\,{\rm cm}\,\gg R_{\star}$, this is not inconsistent with the much smaller blackbody radii inferred from QPE luminosities and temperatures $\sim R_{\odot}$ because the radiation is not in thermal equilibrium with the gas.  The possibility of a larger emission surface owing to an optically thin bremsstrahlung has been considered also in \cite{Krolik&Linial22}.

As in supernovae, the light curve in individual spectral bands will become broader and peak later at longer wavelengths as the diffusion surface where $\tau = c/v$ gradually moves inwards to layers with lower velocity $v < v_{\rm ej}$ and associated longer expansion times (e.g., see \citealt{Kasen+07}; their Fig.~1), broadly consistent with the properties of QPE flares, such as their non-symmetric hardness evolution (e.g., \citealt{Arcodia+21,Arcodia+22}).  Indeed, QPEs measured in harder energy bands are stronger, peak earlier, and have shorter duration compared to when measured at softer bands (e.g., \citealt{Miniutti+19}).\footnote{While supernovae usually evolve monotonically from blue to red, the flare light curves of eRO-QPE1 begin to rise first in lower-energy X-ray bands before evolving to lower energies again after peak \citep{Arcodia+22}; this behavior has not yet been clearly observed from either of the long-short QPE sources.  A detailed model of shock break-out from the disk surface is needed to make predictions for this earliest light curve phase.}

The precise light curve shape will depend on details such as the velocity distribution of the debris and the vertical density profile of the disk.  Further complicating the latter is the likely presence of an additional layer of gas above the disk midplane, created by the accumulation of mass from the disk-star collision ejecta; we estimate the properties of this ``coronal layer'' in Appendix~\ref{sec:corona} and its (typically modest) effect on the flare light curve.  Hydrodynamical simulations of the shock-disk collision for a range of disk/orbit inclination angles, which include the creation and transport of radiation through the expanding ejecta, are needed to make energy-dependent light curve predictions. 

The above calculations apply to a star with a well-defined solid surface of radius $R_{\star} \sim R_{\odot}$.  If the EMRI is instead a point-like compact object (e.g., a $\sim 1 \, \rm M_\odot$ white dwarf or neutron star, or a stellar-mass black hole of mass $m_{\bullet} \sim 10-100M_{\odot}$), then the effective radius for interaction with the disk is the much smaller Bondi radius, $R_{\rm B} \simeq Gm_{\bullet}/v_{\rm K}^{2} = (m_{\bullet}/M_{\bullet})r_0 \ll R_{\star}$.  Repeating the same calculations as above, we show in Appendix \ref{sec:BHcollision} that the resulting timescale and peak luminosity of the compact object-disk collision transient are substantially lower than for a non-degenerate stellar EMRI, e.g. $L_{\rm QPE,\bullet} < L_{\rm Edd}(m_{\bullet}/M_{\bullet})$ (Eq.~\eqref{eq:LpkBH}; see also \citealt{Ivanov+98,Pihajoki+16}), incompatible with observed QPE flares\footnote{While the luminosity produced by a $1-100 \, \rm M_\odot$ compact object EMRI is too dim compared with the observations, we note that the flare temperature is roughly within the observed range of $100-200 \, \rm eV$ (Eqs.~\ref{eq:TBB}, \ref{eq:Tobs}).}  This excludes black holes, neutron stars, or white dwarfs as the source of the disk collisions, unless $m_\bullet/M_\bullet > 10^{-3}$, i.e. for intermediate mass black holes (IMBH); however, IMBH are disfavored as QPE sources based on rate constraints (see Sec.~\ref{sec:BHcollision2}).  

The five key parameters of the light curve model presented in this section are summarized in Table \ref{tab:params}.  Of these, one is directly measurable ($P_{\rm QPE}$) while another two ($\dot{m}$ and $M_{\bullet}$) can in principle be measured or constrained based on the quiescent disk emission and host galaxy properties.  Combined with the three other observables ($L_{\rm QPE},t_{\rm QPE},T_{\rm obs}$), the problem is therefore constrained.

\subsubsection{Emission from the Exposed Disk Midplane}
\label{sec:hole}

In addition to the expanding clouds of shock-heated gas, another potential source of transient emission is radiation from the exposed hot disk midplane, i.e. the walls of the cavity temporarily cleared out by the passage of the star.  The maximum luminosity of the interior surface of the carved hole (of diameter $\simeq 2 R_{\star}$ and height $\simeq h$) is estimated as
\begin{eqnarray}
L_{\rm c} &=& 4\pi R_{\star} h\sigma T_{\rm c}^{4} \simeq \frac{3\pi}{2}c R_{\star} \Sigma v_{\rm K}^{2}\left(\frac{h}{r}\right)^{2} \nonumber \\
&\approx&  3.8\times 10^{41}\,{\rm erg\,s^{-1}}\,\frac{\dot{m}_{-2}\mathcal{R}_{\star}M_{\bullet,6}}{\alpha_{-1}\mathcal{P}_{\rm QPE,4}},
\label{eq:Lhole0}
\end{eqnarray}
where $T_{\rm c} \approx 37$ eV $\alpha_{-1}^{-1/4}\mathcal{P}_{\rm QPE,4}^{-2/5}$ is the midplane temperature (Eq.~\eqref{eq:Tc}).  Both the luminosity and temperature of the midplane emission are potentially consistent with those of QPE flares, suggesting this component could in principle augment the luminosity of the collision-heated ejecta described in the previous section.

A key question, however, is how long the exposed midplane material will generate high-temperature radiation.  The hole pierced by the star will be sealed, and the disk returned approximately to its original state, on the shorter of two timescales:\\
(a) the timescale for radial shear $t_{\rm shear} = 1/\Omega$,
\be
\frac{t_{\rm shear}}{P_{\rm QPE}} = \frac{1}{\pi} \approx 0.32
\ee
(b) the hole sound-crossing time $T_{\rm c_s} = 2R_{\star}/c_s$,
\be
\frac{t_{\rm c_s}}{P_{\rm QPE}} \equiv \frac{2R_{\star}}{c_{\rm s}P_{\rm orb}} = \frac{1}{\pi}\frac{R_{\star}}{h} \approx 0.10 \frac{\mathcal{R}_{\star}}{\dot{m}_{-1}M_{\bullet,6}},
\label{eq:Tcs}
\ee
where $c_s \simeq (h/r)v_{\rm K}$ is the midplane sound speed of the unperturbed disk.  The minimum of $t_{\rm shear}$ and $T_{c_{\rm s}}$ therefore sets the maximum duty cycle of the midplane emission, which is coincidentally comparable to the duty cycle of the shocked-ejecta transient (Eq.~\eqref{eq:duty}).

However, Eqs.~\eqref{eq:Tc}, \eqref{eq:Lhole0}, represent only maxima on the temperature and luminosity, because after the collision the exposed cavity walls will immediately begin to cool radiatively to temperatures $\ll T_{\rm c}$.  This will occur starting on the diffusion timescale $t_{\rm diff,0} \approx 1/(\rho_{\rm c} \kappa c)$ across the photon mean-free path, where $\rho_{\rm c} \simeq \Sigma/(2h)$ is the midplane density.  Insofar as the latter is extremely small, even compared to the time required for the star to pass through the midplane $t_{\rm cross} \sim 2h/v_{\rm K}$, viz.~
\be
\frac{t_{\rm diff,0}}{t_{\rm cross}} \sim \frac{v_{\rm K}}{c}\frac{1}{\tau_{\rm c}} \approx 2\times 10^{-5} \alpha_{-1}\dot{m}_{-1}\left(\frac{r_0}{100R_{\rm g}}\right)^{-2},
\ee
we conclude that radiation from the temporarily exposed midplane material is unlikely to contribute appreciably to the observed QPE emission.  

Another potential effect on the quiescent disk emission is a temporary drop in its flux due to obscuration by the shock-heated debris cloud covering a significant fraction $\gtrsim 10\%$ of the disk's surface.  Although most of the observed quiescent emission comes from the innermost regions of the disk (far from the collision radius), we encourage searching for changes in the soft X-ray emission during and following the (harder) QPE flares.  Indeed, we speculate in Sec.~\ref{sec:GSN069} about the impact of the flare debris in generating the QPO emission observed between the flares in GSN 069 \citep{Miniutti+23}.

\subsection{Lifetime of QPE Activity}
\label{sec:lifetime}

We now consider the processes that ultimately terminate QPE activity, particularly gas drag-induced orbital decay (leading to potential tidal disruption of the EMRI; \citealt{Linial&Sari22}) and gas stripping/ablation by the cumulative effect of disk-collisions.

\subsubsection{Drag-Induced Orbital Decay}
\label{sec:dragdecay}

Over time, interaction with a gaseous disk can lead to substantial changes in the orbit of the star (e.g., \citealt{Syer+91,MacLeod&Lin20}).  At every collision, a fraction $M_{\rm ej}/M_\star$ of the star's orbital energy is dissipated by accelerating the collided material, implying an orbital evolution timescale\footnote{Lense-Thirring precession could in principle bring the stellar orbit into alignment with the plane of the gaseous disk (typically every few years; Eq.~\eqref{eq:TLT}), increasing the rate of swept-up mass and orbital dissipation by a factor $\sim r/h$; however the low duty cycle spent in this configuration $\sim h/r$ will compensate, resulting in only an order-unity change to $\tau_{\rm decay}$.} 
\be
\tau_{\rm decay} = \frac{P_{\rm orb}}{|\dot{T}_{\rm orb}|} \approx \frac{P_{\rm orb}}{2} \frac{M_\star}{M_{\rm ej}} \approx 2000\,{\rm yr} \; \frac{\alpha_{-1}\dot{m}_{-1} M_{\bullet,6} \mathcal{M}_{\star}}{\mathcal{R}_{\star}^{2}},
\label{eq:taudecay}
\ee
where we have used Eq.~\eqref{eq:Mej} for $M_{\rm ej}$. The above expression is valid as long as the orbital eccentricity is not too large, say $e\lesssim 0.5$.  

In Appendix \ref{sec:eccentricity_evolution} we show that the effect of multiple interactions between the star and disk of radial density profile $\Sigma \propto r^{3/2}$ (Eq.~\eqref{eq:tauc}) is to damp an initially mild eccentricity $e \lesssim 0.1.$  As a consequence, the mild eccentricity $e \lesssim 0.1$ responsible for the observed long/short alternating recurrence times, must reflect the residual eccentricity of the EMRI, before it started interacting with the gas disk (the residual eccentricity of a stellar EMRI is indeed typically $\sim 10\%$; \citealt{Linial&Sari22}). This sets an additional constraint on the lifetime of the system as a QPE of the alternating long-short variety, as this eccentricity is expected to dampen over a timescale $\tau_e \approx e \tau_{\rm decay} \lesssim$ centuries.

\subsubsection{Destruction Via Mass Ablation}
\label{sec:ablation}

The star passes through the disk supersonically, experiencing ram pressure on its surface of magnitude
\begin{multline}    \label{eq:Pram}
p_{\rm ram} \simeq \frac{1}{2}\rho_{\rm c} v_{\rm sh}^{2} \simeq \rho_{\rm c}v_{\rm K}^{2} \approx \\ 3.5\times 10^{11}\frac{\mathcal{P}_{\rm QPE,4}^{1/3}}{\alpha_{-1}M_{\bullet,6}^{4/3} \dot{m}_{-1}^{2}} \,{\rm erg\,cm^{-3}},
\end{multline}
where $\rho_{\rm c} = \Sigma/2h$ is the midplane density.

\citet{Liu+15} perform high-resolution hydrodynamical simulations of the collision between the ejecta of a supernova explosion of a binary companion star, in order to calculate the amount of mass stripped from the companion.  Although the present application is different, the physical set-up (a sustained supersonic flow past a star) is sufficiently similar that we can make use of their results.  Based on their simulation results, performed for a range of different binary separations $a_{\rm bin}$, stellar parameters ($M_{\star} = 0.9, 2.7 M_{\odot}$), and explosion energies $E = M_{\rm ej} v_{\rm ej}^{2}/2$, where here $M_{\rm ej}$ and $v_{\rm ej}$ are the mass and mean velocity, respectively, of the supernova ejecta, we find that the stellar mass stripped/ablated per collision is proportional to the ejecta ram pressure ($p_{\rm ram} \propto E/a_{\rm bin}^{3}$; see \citealt{Liu+15}, their Figs.~6, 7) and can be expressed as
\begin{align}     
\label{eq:Liu}
    \frac{\Delta M_{\star}}{M_{\star}} & \sim 10^{-3} \pfrac{p_{\rm ram}}{p_\star} \nonumber \\
   & \approx 5\times 10^{-7}\frac{\mathcal{R}_\star^4 \mathcal{P}_{\rm QPE,4}^{1/3}}{\alpha_{-1} \mathcal{M}_\star^2 M_{\bullet,6}^{4/3}\dot{m}_{-1}^{2}} \,.
\end{align}
where $p_\star \approx GM_\star^2/(4\pi R_\star^4)$ is the star's mean internal pressure. \citet{Armitage+96} performed hydrodynamical simulations of the collision between a red giant and an accretion disk; although the normalization they find differs from that \citet{Liu+15} obtain for a main-sequence star, their results for the dependence of the mass-loss on the stellar velocity, and the mass and thickness of the disk, are all consistent with the above scaling $\Delta M_{\star} \propto p_{\rm ram}$.
 
The lifetime of a solar-type star, before being destroyed via ablation,
\be
\tau_{\rm abl} \approx 
\frac{M_{\star}}{\Delta M_{\star}}\frac{P_{\rm orb}}{2} \approx 880\,{\rm yr}\, \frac{\alpha_{-1}M_{\bullet,6}^{4/3}\dot{m}_{-1}^{2}\mathcal{P}_{\rm QPE,4}^{2/3} \mathcal{M}_\star^2}{\mathcal{R}_\star^4},
\label{eq:tauabl}
\ee
is typically shorter than the lifetime due to orbital decay (Eq.~\ref{eq:taudecay}).

Finally, we note that the quantity of stripped/ablated mass per disk collision can be comparable to the mass of the quiescent disk intercepted/ejected by the star, $M_{\rm ej}$ (Eq.~\eqref{eq:Mej}); this suggests that stripped material could in principle augment the mass and kinetic energy of the collision ejecta entering our estimates in Sec.~\ref{sec:lightcurve}.  

\section{Tidal Disruption Events as QPE Quiescent Disks}
\label{sec:TDE}

Our estimates in Sec.~\ref{sec:lifetime} show that the lifetimes of stars on orbits capable of explaining QPE timescales are typically at most decades to centuries, which is much shorter than the migration time of the stellar EMRI to the galactic nucleus through gravitational wave emission or gas drag.  This disfavors a long-lived AGN disk as the source of gas feeding the SMBH and quiescent X-ray emission from QPE sources, consistent with the lack of a radially-extended AGN emission region \citep{Miniutti+19,Arcodia+21}.  A more promising source of gas is a transient one, such as that created from the tidal disruption of a star.     

\subsection{Stellar Tidal Disruption}

A star of mass $M_{\star}^{\rm 2nd} = \mathcal{M}_{\star}^{\rm 2nd}M_{\odot}$ and radius $R_{\star}^{\rm 2nd}$ is tidally disrupted if the pericenter radius of its orbit, $r_{\rm p}$, becomes less than the tidal radius $r_{\rm T}$ (Eq.~\eqref{eq:Rt}) (e.g., \citealt{Hills75,Rees88}), where the superscript ``2nd'' distinguishes the disrupted star from the already-present EMRI.  The orbital penetration factor is defined as $\beta \equiv r_{\rm T}/r_{\rm p} > 1$.  The most tightly bound stellar debris falls back to the SMBH on the characteristic fall-back timescale (e.g., \citealt{Stone+13,Guillochon&RamirezRuiz13}),
\be
 t_{\rm fb} \simeq 58\,{\rm d}\,M_{\bullet,6}^{1/2}(\mathcal{M}_{\star}^{\rm 2nd})^{1/5},
\label{eq:tfb}
\ee
where the prefactor we have chosen corresponds to the $\beta=1$ disruption for a $\gamma = 5/3$ polytropic star and we have assumed a mass-radius relationship $R_{\star}^{\rm 2nd} \approx (\mathcal{M}_{\star}^{\rm 2nd})^{4/5}R_{\odot}$ appropriate to lower main sequence stars (\citealt{Kippenhahn&Weigert90}).  The resulting rate of mass fall-back at time $t \gg t_{\rm fb}$, expressed in Eddington units, is given by
\begin{multline}
\dot{m}_{\rm fb} \approx \frac{M_{\star}^{\rm 2nd}}{3t_{\rm fb}\dot{M}_{\rm Edd}}\left(\frac{t}{t_{\rm fb}}\right)^{-5/3} \\
\approx 80 \,M_{\bullet,6}^{-3/2}(\mathcal{M}_{\star}^{\rm 2nd})^{4/5}\left(\frac{t}{t_{\rm fb}}\right)^{-5/3}.
\label{eq:Mdotfb}
\end{multline}
Although the process of debris circularization and disk formation at early times may be complex and possibly delayed (e.g., \citealt{Guillochon&RamirezRuiz15,Bonnerot&Lu20,Metzger22b}), there is good evidence that compact disk formation eventually occurs based on late-time X-ray and UV observations of TDEs (e.g., \citealt{Auchettl+17,vanVelzen+19,Jonker+20}). Based on a sample of optical/UV-selected TDE flares observed $5-10$ years after disruption, \citet{vanVelzen+19} detected UV emission consistent with being thermal emission from thin compact disks of radii $\lesssim 3\times 10^{13}$ cm $\sim r_{\rm cir}$.   Thus, we can expect a quasi-steady accretion flow with a slowly-declining rate $\dot{m}_{\rm fb}$ on radial scales of the circularization radius 
\be
r_{\rm cir} \simeq 2r_{\rm T}/\beta.
\label{eq:Rc}
\ee

\subsection{EMRI Destruction}
\label{sec:destruction}

Our results in Sec.~\ref{sec:lifetime} show that stellar EMRIs are likely to lose a significant fraction of their mass via gas ablation before drag-induced orbital decay.  Interestingly, the accretion-rate dependence of the gas ablation destruction time $\tau_{\rm dest} \propto \dot{m}^{2}$ (Eq.~\eqref{eq:tauabl}) suggests that the phase of the TDE leading to the greatest mass-loss is not the initial phase when the mass accretion rate is highest, but later as $\dot{m} \propto t^{-5/3}$ drops and the midplane density $\rho_{\rm c} \propto \dot{m}/h^{3} \propto \dot{m}^{-2}$ and corresponding ram pressure $\propto \rho_{\rm c}$ on the star, increases.  

Inserting the mass-fall back rate (Eq.~\eqref{eq:Mdotfb}) into Eq.~\eqref{eq:tauabl}, we can estimate when the EMRI will lose a significant fraction of its mass by setting $\tau_{\rm abl} = t$.  This occurs at an accretion rate
\be
\dot{m}_{\rm dest} \approx 1.3\times 10^{-2}\frac{ \mathcal{R}_\star^{20/13} (\mathcal{M}_{\star}^{\rm 2nd})^{17/65}}{\alpha_{-1}^{5/13} \mathcal{M}_\star^{10/13} M_{\bullet,6}^{2/3}\mathcal{P}_{\rm QPE,4}^{10/39}} \, ,
\label{eq:mdotdest}
\ee
as achieved on a timescale after the TDE,
\be
\tau_{\rm dest} 
\approx 24.7\,{\rm yr}\; \frac{\alpha_{-1}^{3/13}(\mathcal{M}_{\star}^{\rm 2nd})^{34/65} \mathcal{M}_\star^{6/13} \mathcal{P}_{\rm QPE,4}^{2/13}}{\mathcal{R}_\star^{12/13}} \,.
\label{eq:taudest}
\ee 
This implies the EMRI will lose a significant fraction of its mass, either leading to removal of its envelope or complete destruction, on a timescale of roughly a decade.  The characteristic accretion rates $\dot{m}_{\rm dest} \sim 10^{-2}-10^{-1}$ over which a star will spend the most time generating QPEs before significant mass-loss, notably coincides with those corresponding to radiatively-efficient thin-disks, consistent with the disk model adopted throughout this paper (Sec.~\ref{sec:disk}) and the evolution of the luminosity/temperature of the quiescent emission in GSN 069 over $\gtrsim 12$ years of observations (\citealt{Miniutti+23}).

In fact, the mass-input rate to the disk from stellar ablation $\dot{M}_{\rm abl} = \Delta M_{\star}/P_{\rm QPE}$ (Eq.~\eqref{eq:Liu}) can become comparable to that from the TDE fallback $\dot{M}_{\rm fb}$, on a somewhat earlier timescale,
\be
\tau_{\dot{M}} \approx 10.3\,{\rm yr}\,\alpha_{-1}^{1/5}M_{\bullet,6}^{1/15}\mathcal{M}_{\star}^{5/3}\mathcal{R}_{\star}^{-12/5}(\mathcal{M}_{\star}^{\rm 2nd})^{1/5}\mathcal{T}_{\rm QPE}^{2/15}.
\label{eq:tauMdot}
\ee
Thus, a flattening of the quiescent X-ray light-curve relative to the canonical $\propto t^{-5/3}$ decay may occur on a timescale $\tau_{\dot{M}}$, potentially also changing the destruction rate compared to the estimate (Eq.~\eqref{eq:taudest}) which neglects this additional source of mass-input to the disk. If the mass accretion rate through the disk near the collision radius were to decay differently from the canonical $\propto t^{-5/3}$ fall-back rate (e.g., due to delayed viscous accretion, $\dot{M} \propto t^{-1.2}$; \citealt{Cannizzo+90,Shen&Matzner14,Auchettl+17}), the normalization and parameter dependencies of Eq.~\eqref{eq:tauMdot} would be moderately different.

If we instead assume that the disk reaches a regulated state in which the accretion rate is dominated by mass-input from stellar ablation instead of TDE fall-back (or whatever gas accretion event triggered stellar mass-loss to begin with), i.e. $\dot{M} = \dot{M}_{\rm abl}$, this defines a second characteristic accretion rate
\be
\dot{m}_{\rm dest,2} \approx 0.074\mathcal{M}_{\star}^{-5/3}\mathcal{R}_{\star}^{4}\alpha_{-1}^{-1/3}M_{\bullet,6}^{-4/9}\mathcal{P}_{\rm QPE,4}^{-2/9},
\ee
and corresponding destruction time
\begin{eqnarray}
\tau_{\rm dest,2} &\approx& \frac{M_{\star}}{\dot{m}_{\rm destr,2}\dot{M}_{\rm Edd}} \nonumber \\
&\approx& 480\,{\rm yr}\,\mathcal{M}_{\star}^{8/3}\mathcal{R}_{\star}^{-4}\alpha_{-1}^{1/3}M_{\bullet,6}^{-5/9}\mathcal{P}_{\rm QPE,4}^{2/9}.
\label{eq:taudest2}
\end{eqnarray}
For $\dot{m}_{\rm dest,2} \gtrsim \dot{m}_{\rm dest}$, the disk may thus enter a regulated state with $\dot{m} \approx \dot{m}_{\rm dest,2}$ with a flat quiescent light curve and corresponding longer destruction time $\tau_{\rm dest,2} \gg \tau_{\rm dest}$ (we shall return to implications of the $M_{\bullet}-$dependence of $\tau_{\rm dest,2}$ in Sec.~\ref{sec:why}).  

\subsection{QPE Rates: Most EMRIs will Experience a TDE}
\label{sec:rates}

A quasi-circular EMRI on an orbit of semi-major axis $a \sim r_{0} \sim r_{\rm circ}$ comparable to the circularization radius of the TDE disk will undergo gravitational wave-driven inspiral on a timescale
\begin{multline}
\tau_{\rm GW} \equiv \frac{a}{\dot{a}_{\rm GW}} \simeq \frac{5}{64}\frac{c^{5}a^{4}}{G^{3}M_{\star}M_{\bullet}^{2}} \simeq \\
8\times 10^{5}\,{\rm yr}\,\frac{\mathcal{R}_{\star}^{4}}{\mathcal{M}_{\star}^{7/3}M_{\bullet,6}^{2/3}}\left(\frac{a}{r_{\rm cir}}\right)^{4},
\label{eq:tauGW}
\end{multline}
where in the final equality we have used Eq.~\eqref{eq:Rc} for $\beta = 1$ and have assumed for simplicity that both the EMRI, and the tidally-disrupted star, possess the same masses and radii.  The fact that this timescale is generally longer than the average interval between consecutive TDE in a typical galactic nucleus ($T_{\rm TDE} \sim 1/\dot{N}_{\rm TDE} \sim 10^{4}-10^{5}$ yr; e.g., \citealt{Magorrian&Tremaine99,Stone&Metzger16,Yao+23}) shows that most EMRIs will experience at least one TDE event before migrating inwards sufficiently past the circularization radius to undergo Roche lobe overflow onto the SMBH. 

Stellar EMRIs are produced by single-single scattering, or the Hills mechanism \citep{Hills88}, at respective rates \citep{Linial&Sari22}
\begin{align}
    \dot{N}_{\rm EMRI,single} \approx 10^{-7} \; M_{\bullet,6}^{1.1} \, \; \rm yr^{-1} \, \\
    \dot{N}_{\rm EMRI,Hills} \approx 10^{-5} \; \pfrac{f_b}{0.1} M_{\bullet,6}^{-0.25} \; \rm yr^{-1} \,.
    \label{eq:NdotEMRI}
\end{align}
where $f_b$ is the fraction of sufficiently tight binaries within the SMBH's radius of influence that contribute to the EMRI production through the Hills mechanism.

Assuming that QPEs are detectable for a timescale comparable to the estimated EMRI destruction time $\tau_{\rm dest} \sim 10$ yr (Eq.~\eqref{eq:taudest}), a lower limit to the fraction of galaxies predicted to host a QPE is given by
\be
f_{\rm QPE} \simeq \dot{N}_{\rm EMRI}\tau_{\rm dest} = 10^{-5}\left(\frac{\dot{N}_{\rm EMRI}}{10^{-6}\,\rm yr^{-1}}\right)\left(\frac{\tau_{\rm dest}}{10\,\rm yr}\right).
\label{eq:fQPE1}
\ee
Note that the above calculation accounts for the fact that a significant fraction of stellar EMRIs eventually evolve to tight, mildly eccentric orbits capable of producing QPEs, and thus $\dot{N}_{\rm EMRI}$ comprises the replenishment rate of QPEs, of assumed active duration $\tau_{\rm dest}$. On the other hand, if EMRIs can survive more than one TDE before being destroyed ($N_{\rm TDE}> 1$) then the expected rate will be larger by at least a factor $\propto N_{\rm TDE}$. Insofar that the occurrence of a TDE is necessary to create a compact gaseous disk, the maximum QPE occupation fraction is then set by the TDE rate,
\be
f_{\rm QPE} \simeq \dot{N}_{\rm TDE}\tau_{\rm dest} = 10^{-4}\left(\frac{\dot{N}_{\rm TDE}}{10^{-5}\,\rm yr^{-1}}\right)\left(\frac{\tau_{\rm dest}}{10\,\rm yr}\right)
\label{eq:fQPE2}
\ee
These ranges are broadly consistent with the observed QPE occurrence fraction $f_{\rm QPE} \sim 10^{-5}$ (R.~Arcodia, private communication).\footnote{As discussed in \citet{Metzger+22}, this rate can be very roughly estimated by dividing the 2 sources discovered with eROSITA \citep{Arcodia+21} by the product of the co-moving volume out to the most distant source eRO-QPE1 at redshift $z = 0.0505$ of $\approx 0.04$ Gpc$^{-3}$ by the density of Milky Way–like galaxies of $6\times 10^{6}$ Gpc$^{-3}$ as a proxy for potential QPE hosts.  Since eRO-QPE2 was three times closer than eRO-QPE1, if the QPEs which exhibit regular ``long-short'' behavior such as eRO-QPE2 indeed form a distinct class, then their implied rate would be lower.}  

Our calculations in Sec.~\ref{sec:destruction} consider the destruction of an EMRI on a quasi-circular orbit close to the tidal radius as needed to explain observed QPEs in our scenario.  However, most EMRIs arrive to the galactic nucleus as the result of gravitational wave-driven inspiral starting from orbits with much larger semi-major axis and higher eccentricity.  It is thus likely that migrating EMRIs must pass through the gaseous disks of many TDE disks on their way to becoming circular EMRIs, bringing into question whether they could survive long enough to become quasi-circular QPE sources with short orbital periods.  

Insofar that the total stellar mass-loss (Eq.~\eqref{eq:Liu}) depends on the total number of midplane passages, it is thus relevant to consider whether the total number of such disk-star collisions during the inward migration of the EMRI is dominated by the time the star spends at larger or small semi-major axis.  The number of star-disk collisions at each semi-major axis $a$ can be crudely estimated as,
\be N_{\rm coll}(a) = \left(\frac{\Delta T_{\rm TDE}}{T_{\rm TDE}}\right)\frac{\tau_{\rm GW}(a,r_p)}{P_{\rm orb}(a)} \propto a^{-1},
\ee
where $T_{\rm TDE} \approx \dot{N}_{\rm TDE}^{-1} \approx 10^4-10^5 \, \rm yr$ is the average interval between TDE and $\Delta T_{\rm TDE}$ is the average lifetime of the TDE disk. $\tau_{\rm GW}(a,r_p) \propto a^{1/2}$ is the evolution timescale of an orbit of semi-major axis $a$ and pericenter distance $r_p$ due to gravitational wave emission (we assume $r_p$ is essentially fixed during the GW circularization). Thus, the total mass-loss from the star is dominated by small radii, consistent with EMRI destruction occurring only after the stellar orbit has largely circularized near radii $R_{\rm circ} \sim r_{\rm T}$ corresponding to observed QPE. 

\subsection{Conditions to Observe a QPE}
\label{sec:why}

It is useful to compare the predicted QPE luminosities and temperatures from Sec.~\ref{sec:lightcurve} to those of the quiescent disk emission (Sec.~\ref{sec:disk}).  Using Eqs.~\eqref{eq:Lpk}, \eqref{eq:LQ} and  Eqs.~\eqref{eq:Tobs}, \eqref{eq:TQ}, we find that
\be
\frac{L_{\rm QPE}}{L_{\rm Q}} \approx 0.02\mathcal{R}_{\star}^{2/3}\dot{m}_{-1}^{-2/3}\mathcal{P}_{\rm QPE,4}^{-2/3};
\label{eq:Lratio}
\ee
\begin{eqnarray}    
    \frac{T_{\rm obs}}{T_{\rm Q}} 
    \approx 1.3\frac{\alpha_{-1}^{5/2}\dot{m}_{-1}^{11/4}M_{\bullet,6}^{14/3}}{\mathcal{R}_{\star}^{1/3}\mathcal{P}_{\rm QPE,4}^{13/3}} \,.
    \label{eq:TQratio}
\end{eqnarray}
Even though the bolometric luminosity of the collision-flare is generally less than the quiescent disk, the higher temperature of the flare emission can still render it detectable, given the increasing sensitivity of X-ray telescopes at higher photon energies $\gtrsim 100$ eV.  The ratio $T_{\rm obs}/T_{\rm Q}$ is thus a rough proxy for the observability of QPE emission, as illustrated with a few models in the bottom right panel of Fig.~\ref{fig:LpkTobs}.

The steep dependence $T_{\rm obs}/T_{\rm Q} \propto P_{\rm QPE}^{-13/3}$ implies that bursts with shorter $P_{\rm QPE} \sim P_{\rm QPE,min}$ are more easily detectable than those with longer $P_{\rm QPE}$.  This may contribute to the absence of QPE sources with much longer orbital periods than the observed sample.  This is despite the likelihood that the same EMRIs which generate short-period QPEs today may have collided with the gaseous disks of many TDEs as they migrated into the nucleus from larger distances (Sec.~\ref{sec:rates}).  

On the other hand, the seeming preference for QPE detection with increasing SMBH mass is at odds with the low-mass galaxies of QPE systems (e.g., \citealt{Wevers+22}) and does not find an obvious explanation in our model.  Both the rate of stellar EMRIs (Eq.~\eqref{eq:NdotEMRI}) and TDEs (e.g., \citealt{Stone&Metzger16}) are predicted to rise towards lower SMBH mass, but only relatively gradually $\propto M_{\bullet}^{-1/4}$ \citep[e.g.,][]{Linial&Sari22}. The stellar-EMRI production rate is typically dominated by the Hill's mechanism, occurring when tidally split binaries leave behind a bound star on a highly eccentric orbit of semi-major axis $a_{\rm Hills}$ \citep{Linial&Sari22}. Since the overall number of stars residing within the SMBH's radius of influence is smaller in low mass systems, the average number of stars orbiting the SMBH at similar radii $\sim a_{\rm Hills}$ may be less than unity, implying that captured stars evolve primarily by gravitational wave emission rather than stochastic angular momentum diffusion, enhancing the stellar-EMRI rate and consequentially, the QPE rate, in low $M_\bullet$ systems. The critical radius below which the average number of stars assuming a Bahcall-Wolf density profile drops below unity is roughly $r_1 \approx 5\times 10^{13} \, M_{\bullet,6}^{-0.3}$ cm, while for a near-contact binary, $a_{\rm Hills} \approx 2\times 10^{15}\, M_{\bullet,6}^{2/3}$ cm. Therefore, the ratio $a_{\rm Hills}/r_1 \propto M_\bullet$, and falls below unity for $M_\bullet \lesssim 3\times 10^4 \; \rm M_\odot$. In such low mass systems, two-body scatterings are expected to become inefficient relative to GW inspiral. A similar qualitative argument was invoked by \cite{Lu&Quataert22}.

Another speculative explanation for the low-$M_{\bullet}$ QPE preference comes from our estimate in Sec.~\ref{sec:ablation} that the EMRI destruction time, and hence the observed QPE occupation fraction $f_{\rm QPE} \propto \tau_{\rm dest}$ (Eq.~\eqref{eq:fQPE1}, \eqref{eq:fQPE2}), could grow with decreasing $M_{\bullet}$ (see Eq.~\eqref{eq:taudest2} and surrounding discussion).

\section{Applications}
\label{sec:applications}

\subsection{Long-Short Alternating QPE}

\subsubsection{GSN 069}
\label{sec:GSN069}

One of the best-studied QPE sources is GSN 069 \citep{Saxton+11,Miniutti+13}, which generated QPE outbursts over observations spanning at least one year \citep{Miniutti+19,Miniutti+23}.  The quiescent X-ray-emitting accretion flow was likely generated by a TDE or TDE-like flare \citep{Shu+18,Sheng+21}, which was followed$-$immediately after the final epoch QPE activity was detectable$-$by a second TDE-like flare of similar amplitude \citep{Miniutti+23}. 

If the observed quiescent luminosity $L_{\rm Q} \sim 10^{43}$ erg s$^{-1}$ and temperature $T_{\rm Q} \approx 50$ eV from GSN 069 is thermal disk emission (Eqs.~\eqref{eq:LQ}, \eqref{eq:TQ}), then for SMBH masses $M_{\bullet} \sim 3\times 10^{5}-3\times 10^{6}M_{\odot}$ consistent with the host galaxy \citep{Miniutti+23} we infer a range of Eddington accretion fraction $\dot{m} \sim 10^{-2}-0.3$.  As summarized in Fig.~\ref{fig:LpkTobs}, given the observed QPE period $P_{\rm QPE} \approx 9$ hr, our light curve model (Sec.~\ref{sec:lightcurve}; Eqs.~\eqref{eq:tpk}, \eqref{eq:Lpk}, \eqref{eq:Tobs}) predicts QPE flare luminosities $L_{\rm QPE} \approx (1-5)\times 10^{41}\,{\rm erg\,s^{-1}}\mathcal{R}_{\star}^{2/3}$, durations $t_{\rm QPE} \sim (0.1-0.5)\mathcal{R}_{\star}$ hr.  These are in reasonable agreement with those observed ($L_{\rm QPE} \approx 10^{42}$ erg s$^{-1}$; $t_{\rm QPE} \sim 1$ hr; Table \ref{tab:sources}) for a star of a couple solar radii orbiting a SMBH of mass $\sim (1-3)\times 10^{6}M_{\odot}$.  The observed temperature $T_{\rm obs} \sim 100$ eV can can also be reproduced by the model (Eq.~\eqref{eq:Tobs}; Fig.~\ref{fig:LpkTobs}), though there is substantial degeneracy between the SMBH mass, accretion rate, and disk viscosity $\alpha.$

Although we do not want to over-interpret precise numbers given the simplifying assumptions of our analytic model, we note that a radius slightly larger than a solar radius does not necessarily require a massive star; the mass-loss rate of the star from disk-star interactions $\sim 0.1M_{\odot}$ yr$^{-1}$ is sufficiently rapid for the star's radius to be significantly inflated relative to a main-sequence star in thermal equilibrium (e.g., \citealt{Linial&Sari17}).

The existence of a long-term modulation in the QPE signal intensity and recurrence are also mentioned in \citet{Miniutti+23}; we speculate these to be related to orbital evolution effects, such as general relativistic apsidal precession, which for the system properties of GSN 069 will take place over a timescale $T_{\epsilon} \sim$ weeks to months (Eq.~\eqref{eq:Tepsilon}).

QPE emission was not detected in GSN 069 in 2015, a delay of at least several years after the creation of the gaseous TDE disk. Although the ratio of the flare to quiescent emission temperature $T_{\rm obs}/T_{\rm Q} \propto \dot{m}^{17/12}$ (Eq.~\eqref{eq:TQratio}) is higher when $\dot{m}$ is higher earlier in the TDE, the luminosity contrast $L_{\rm QPE}/L_{\rm Q} \propto \dot{m}^{-2/3}$ (Eq.~\eqref{eq:Lratio}) will be lower.  A more careful analysis of the X-ray data within the context of a two-component (quiescent disk + flare) model would be required to determine if the early non-detection of QPE emission is constraining.  An alternative explanation for the delayed onset of QPE emission is that the orbit of the EMRI resides outside the circularization radius of the TDE, and the disk takes several years to form and viscously spread outwards to meet the EMRI orbit (G. Miniutti, private communication).  

 \citet{Miniutti+23} found evidence for quasi-periodic oscillations (QPO) in the timing properties of the quiescent X-ray emission phases between each QPE flare.\footnote{\citet{Song+20} also claimed evidence for QPO emission from the QPE source RX J1301.9+2747 \citep{Giustini+20}.}  Insofar that the total amount of mass excavated by disk-star collisions can be significant compared that flowing inwards through the disk (Appendix~\ref{sec:corona}), it would not be surprising for the quiescent flow to be significantly perturbed by each collision (e.g., \citealt{Sukova+21}).  Furthermore, debris ejected from the collision site will expand at velocity $\sim v_{\rm K}$ in all directions, including towards the SMBH and innermost regions of the accretion flow.  This could significantly perturb the quiescent disk emission on a timescale as short as $\delta t_{\rm Q} \sim r_0/v_{\rm K} \approx P_{\rm QPE}/\pi$; indeed, for GSN 069 ($P_{\rm QPE} = 9$ hr) this gives $\delta t_{\rm Q} \approx 2.8$ hr, precisely the observed delay when the QPO emission peaks after each flare \citep{Miniutti+23}.

Gas input to the disk from star-disk collision will become comparable to that due to fall-back accretion from the TDE on a timescale $\tau_{\dot{M}}$ of several years (Eq.~\eqref{eq:tauMdot}), taking $\mathcal{M}_{\star}^{\rm 2nd} \sim 0.5M_{\odot}$ for the mass of the disrupted star (based on the integrating the radiated X-ray energy assuming radiatively efficient accretion; \citealt{Miniutti+23}).  Indeed, the decay of the quiescent X-ray light curve of GSN 069 was notably gradual compared to those typical of X-ray TDEs \citep{Shu+18}.  

 Removal of a significant fraction of the star's mass due to disk-star collisions will occur on a timescale $\tau_{\rm dest} \sim $ 10 years after the TDE (Eq.~\eqref{eq:taudest}), roughly consistent with when the QPE activity was observed to cease.  Insofar that the second TDE-like flare began to rise almost at this very same time \citep{Miniutti+23}, we speculate that the second ``TDE" was powered not by disruption of yet another star, or the remaining core of the first TDE \citep{Miniutti+23}, but rather by the rapid disruption and accretion of the QPE-generating EMRI following a sudden acceleration in its mass-loss rate.  Such a runaway could arise from strong positive feedback between the mass-loss-driven expansion of the star and the mass-loss per collision ($M_{\rm ej} \propto R_{\star}^{4}M_{\star}^{-2}$; Eq.~\ref{eq:Liu}).  Alternatively, this same expansion may lead to Roche lobe overflow and unstable mass transfer onto the SMBH \citep{Linial&Sari22}; indeed, the stellar radius compatible with the light curve properties is close to the Roche radius size $R_{\rm RL} \approx 3.4R_{\odot} \mathcal{M}_{\star}^{1/3}(P_{\rm QPE}/9\,{\rm hr})^{2/3}$ (Eq.~\eqref{eq:TQPEmin}).  The disappearance in the final XMM epoch before the final TDE of regular QPE recurrence properties exhibited in the earlier epochs \citep{Miniutti+23} could support significant evolution of either the stellar or orbital properties.  If the EMRI is an evolved star, the dynamical stripping of its outer envelope does not preclude the longer-term survival of a core, in which case QPE emission from disk-star collisions might resume following the second TDE.    

 The accretion-powered X-ray transient generated by mass stripped from a star on a mildly eccentric or circular orbit will likely differ from that generated in the usual case of a near-parabolic TDE (e.g., \citealt{Xin+23}).   While in a normal TDE the peak emission time is generally controlled by the rate at which the marginally-bound debris falls back to the SMBH, in the circular case the matter already begins tightly bound with a short orbital period and the flare duration is instead likely to be set by the viscous timescale of the disk or the timescale for the runaway mass-loss (which we have shown itself likely depends on the time-evolving $\dot{M}-$dependent properties of the gaseous disk).  We defer an exploration of disk creation and associated X-ray emission from runaway EMRI mass-loss events, and its consistency with the X-ray outburst(s) observed from GSN 069, to future work.

\subsubsection{eRO-QPE2}
\label{sec:eROQPE2}

The measured 0.5-2 keV band quiescent luminosity $\sim 10^{41}$ erg s$^{-1}$ of eRO-QPE2 \citep{Arcodia+21} should be taken as a lower limit on the total quiescent disk luminosity, as the latter is sensitive to the assumed spectral shape and the treatment of absorption.  This translates into a lower limit on the disk accretion rate $\dot{m} \gtrsim 10^{-3}-10^{-2}$ for $M_{\bullet} \sim 10^{5}-10^{6}M_{\odot}$.  Given the observed QPE period $P_{\rm QPE} \approx 2.5$ hr and assuming a solar-type star for simplicity, our light curve model (Sec.~\ref{sec:lightcurve}) predicts $L_{\rm QPE} \approx 5\times 10^{41} M_{\bullet,6}\dot{m}_{-1}^{1/3}\,{\rm erg\,s^{-1}}$ and $P_{\rm QPE} \approx 0.1M_{\bullet,6}^{-2/3}\dot{m}_{-1}^{-1/2}$ hr.  For $M_{\bullet,6} \sim \dot{m}_{-1} \sim 1$, the predictions and observations match within a factor $\lesssim 2$ (Table \ref{tab:sources}).  

All else being equal, QPE sources with shorter periods like eRO-QPE2 are predicted to generate shorter, more luminous, and harder QPE flares than longer period sources like GSN 069, consistent with the observed trends between the two sources discussed in this section (Fig.~\ref{fig:LpkTobs}).  

\subsection{Generalization to high eccentricity}
\label{sec:highe}
In Sec.~\ref{sec:stardisk} we have assumed that the star follows a mildly eccentric orbit, with $e \approx 0.1$.  Here we consider some generalization of the model to a highly eccentric orbit, with $e\gtrsim 0.5$. In such case, deviations between the timing of consecutive flares become considerable, and the approximation that both star-disk collisions occurring each orbit take place at $r\approx r_0$, are invalid. The timing of the star-disk collisions generally depend on the relative inclination and eccentricity vector of the orbit with respect to the disk. If the orbit is highly eccentric, with pericenter $r_{\rm p} \ll r_0$, collisions may occur over a wide range of radii, from $r_{\rm p}$ to $(1+e)r_0 \approx 2r_0$. 

If the disk extends past the star's apocenter, two collisions will occur per orbit, at radii $\{ r_1,r_2\} $, at intervals $\Delta T_1 \leq \Delta T_2$, such that their sum equals the orbital period $\Delta T_1 + \Delta T_2 = P_{\rm orb}$. When $e$ is relatively small, $(T_1-T_2)/P_{\rm orb} \sim \mathcal{O}(e) \ll 1$, and we have thus assumed $P_{\rm QPE} \approx P_{\rm orb}/2$.
However, for high $e$, the time difference between consecutive collisions can be as short as $\Delta T_1 \approx P_{\rm orb} (r_{\rm p}/r_0)^{3/2} = P_{\rm orb} (1-e)^{3/2}$, and as long as $\Delta T_2 \lesssim P_{\rm orb}$. If both collisions produce observable flares, they will appear to be produced in pairs, spaced with highly asymmetric intervals, $\Delta T_1 \ll \Delta T_2$.

Since at least one of the two collisions occurs at around $r_{\rm p}$, which is considerably smaller than $r_0$, the expressions in Sec.~\ref{sec:stardisk} are not directly applicable to the high $e$ case. For example, the Keplerian velocity near $r_{\rm p}$ is
\begin{equation}
    v_{\rm K}(r_{\rm p}) \approx \pfrac{(1+e)G M_{\bullet}}{r_{\rm p}}^{1/2} \approx \pfrac{GM_\bullet}{r_0}^{1/2} \pfrac{2}{1-e}^{1/2} \,,
\end{equation}
higher by a factor $\sqrt{2/(1-e)}$ compared with Eq.~\eqref{eq:vK}. Previously, we assumed $P_{\rm QPE} \approx P_{\rm orb}/2$, and therefore associated $P_{\rm QPE}$ with $r_0$ (i.e., Eq.~\eqref{eq:TQPE}). We can generalize some of the expressions by defining $P_{\rm QPE} = (\Delta T_1 + \Delta T_2)/2$, and considering
\begin{equation}
    r_{\rm p} \approx r_0 (1-e) \approx 1.4\times 10^{13} \; {\rm cm} \; (1-e) \mathcal{T}_{\rm QPE}^{2/3} M_{\bullet,6}^{1/3} \,.
\end{equation}
such that corrections of order $(1-e)$ raised to different powers are introduced in some of the previous expressions. For example $M_{\rm ej} \to M_{\rm ej} (1-e)^{3/2}$ (Eq.~\eqref{eq:Mej}), $P_{\rm QPE} \to P_{\rm QPE} (1-e)$ (Eq.~\eqref{eq:tpk}), $L_{\rm QPE} \to L_{\rm QPE} (1-e)^{-1}$ (Eq.~\eqref{eq:Lpk}), $\mathcal{D}\to \mathcal{D}(1-e)^{-1/2}$ (Eq.~\eqref{eq:duty}) and $k_{\rm B} T_{\rm obs} \to k_{\rm B} T_{\rm obs} (1-e)^{-9/2}$ (Eq.~\eqref{eq:Tobs}). Apsidal and nodal precession would introduce order unity corrections to these terms, varying on timescales $T_\epsilon \to T_\epsilon(1-e)$ (Eq.~\eqref{eq:Tepsilon}) and $T_\Omega \to T_\Omega (1-e)^{3/2}$ (Eq.~\eqref{eq:TLT}).

Another subtlety introduced when considering high $e$ appears if the disk is confined to a smaller radius $r_{\rm d}$, $r_p \lesssim r_{\rm d} \ll r_0$. In this case, the star might intercept the disk only once per orbit, rather than twice. Flares would then occur periodically, with $P_{\rm QPE} \approx P_{\rm orb}$.

\subsection{Black Hole-Disk Collisions}
\label{sec:BHcollision2}

In Appendix \ref{sec:BHcollision}, we show that the collision of a compact object with an SMBH accretion disk produces flares too dim and too short to explain QPE observations, unless the compact object EMRI is an IMBH of mass $m_{\bullet} \gtrsim 10^{3}M_{\odot}$.  However, aside from their predicted formation rates being far too low, IMBH-SMBH binaries as dominant QPE sources is disfavored based on the implications this would have for SMBH populations.  The minimum average rate of mass growth of a SMBH due to mergers with compact objects of mass $m_{\bullet}$ needed to explain the QPE population can be estimated as
\begin{eqnarray}
\dot{M}_{\bullet} \sim \frac{m_{\bullet}}{\tau_{\rm GW}}f_{\rm QPE} \simeq \frac{64}{5}f_{\rm QPE}\frac{M_{\bullet}}{c/R_{\rm g}}\left(\frac{m_{\bullet}}{M_{\bullet}}\right)^{2}\left(\frac{r_0}{R_{\rm g}}\right)^{-4},
\end{eqnarray}
where $\tau_{\rm GW}$ is the gravitational wave inspiral time through the range of QPE orbital periods (Eq.~\eqref{eq:tauGW} with $m_{\bullet}$ replacing $M_{\star}$ and $a = r_0$) and $f_{\rm QPE} = 10^{-5}$ is the observed occupation fraction of QPE sources in galactic nuclei (Sec.~\ref{sec:rates}).  This is a minimum rate because it only counts mergers that occur when a gaseous disk is present to enable collision-powered QPEs.  Thus, the timescale for the SMBH to appreciably grow
\begin{multline}
\tau_{\rm grow} \sim \frac{M_{\bullet}}{\dot{M}_{\bullet}} \nonumber \\
\approx  4\times 10^{9}\,{\rm yr}\,\left(\frac{f_{\rm QPE}}{10^{-5}}\right)^{-1}\frac{\mathcal{P}_{\rm QPE,4}^{8/3}}{M_{\bullet,6}^{5/3}}\left(\frac{m_{\bullet}}{10^{3}M_{\odot}}\right)^{-2},
\end{multline}
is comparable to or less than the Hubble time for $m_{\bullet} \gtrsim 10^{3}M_{\odot}$.  If true, SMBHs would need to grow to their present masses mostly through black hole mergers, contrary to observations which attribute most growth to gaseous accretion (e.g., \citealt{Soltan82}). Alternatively, if QPEs preferentially occur in a rare SMBH sub-population which indeed grow primarily through IMBH mergers, the relevant $f_{\rm QPE}$ of that sub-population to explain their occurrence rate must greatly exceed $10^{-5}$, implying $\tau_{\rm grow} \ll$ Hubble time, incompatible with the relatively modest SMBH mass of the known QPE hosts.

\subsubsection{Application to OJ 287}
\label{sec:OJ287}

The analysis above obviously does not the exclude the existence of disk-collision transients from IMBH-SMBH or SMBH-SMBH binaries, as long as they are unassociated with the bulk of the QPE sources.  One such candidate is the SMBH binary candidate OJ 287 at redshift $z = 0.306$ (e.g., \citealt{Sillanpaa+88,Valtonen+08,Valtonen+12,Komossa+23}), which exhibits quasi-periodic optical/UV flares that occur in pairs separated by $\sim$1-2 years reoccurring every $\sim$12 years in the observer frame.  Among several models, it was proposed that these flares are powered by the collision between the disk of a SMBH (e.g., \citealt{Lehto&Valtonen96,Valtaoja+00}) and a lighter SMBH on an eccentric orbit $e \approx 0.7$ \citep{Sillanpaa+88}.  Although the mass of the primary(secondary) SMBH was estimated to be $2\times 10^{10}M_{\odot}(1.5\times 10^{8}M_{\odot}$) through modeling the timing of the flares (e.g., \citealt{Valtonen+08}), recent observations by \citet{Komossa+23} favor a substantially lower primary mass $\approx 10^{8}M_{\odot}$.

\begin{figure}
    \centering
    \includegraphics[width=0.45\textwidth]{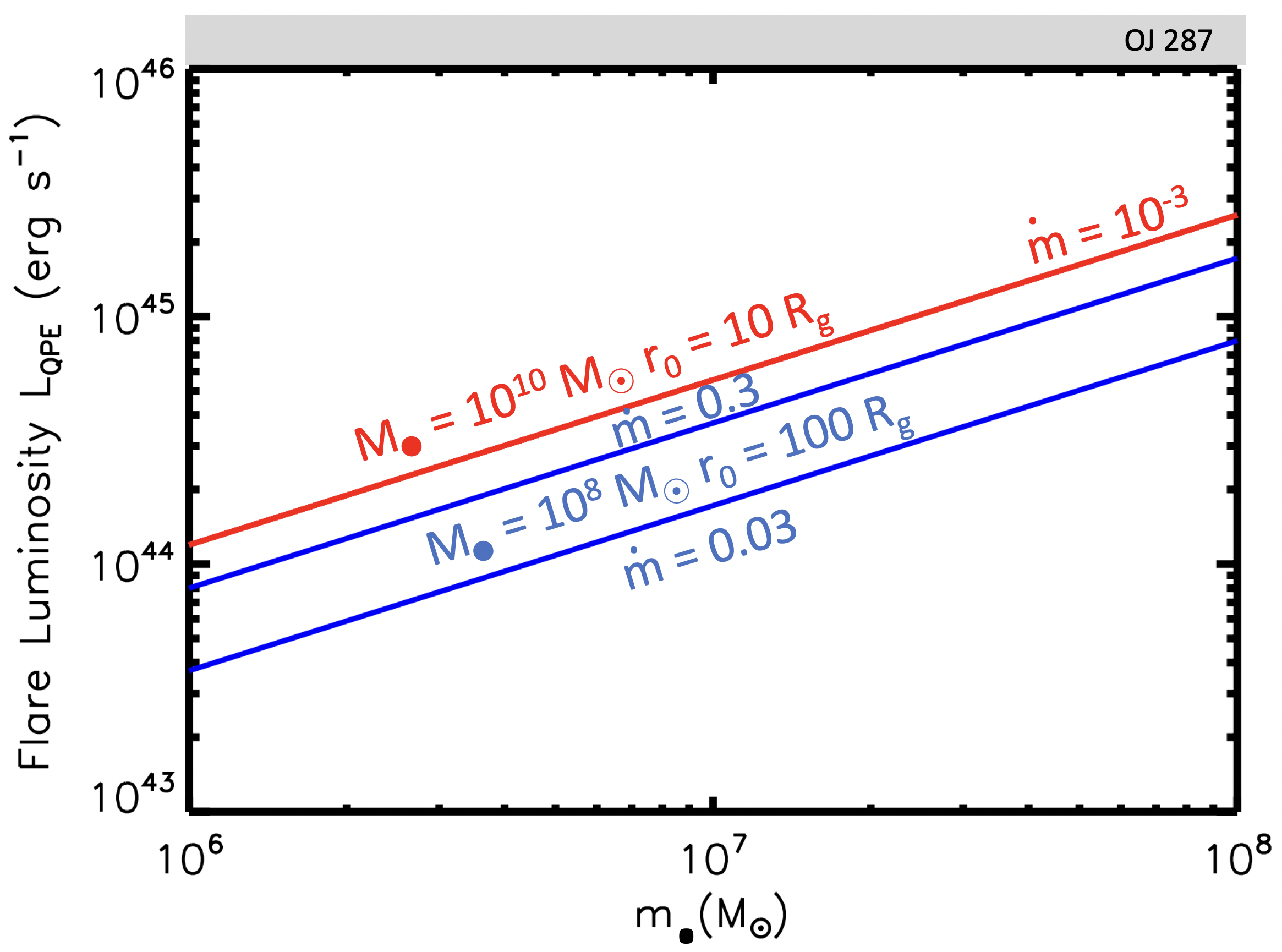}
    \includegraphics[width=0.45\textwidth]{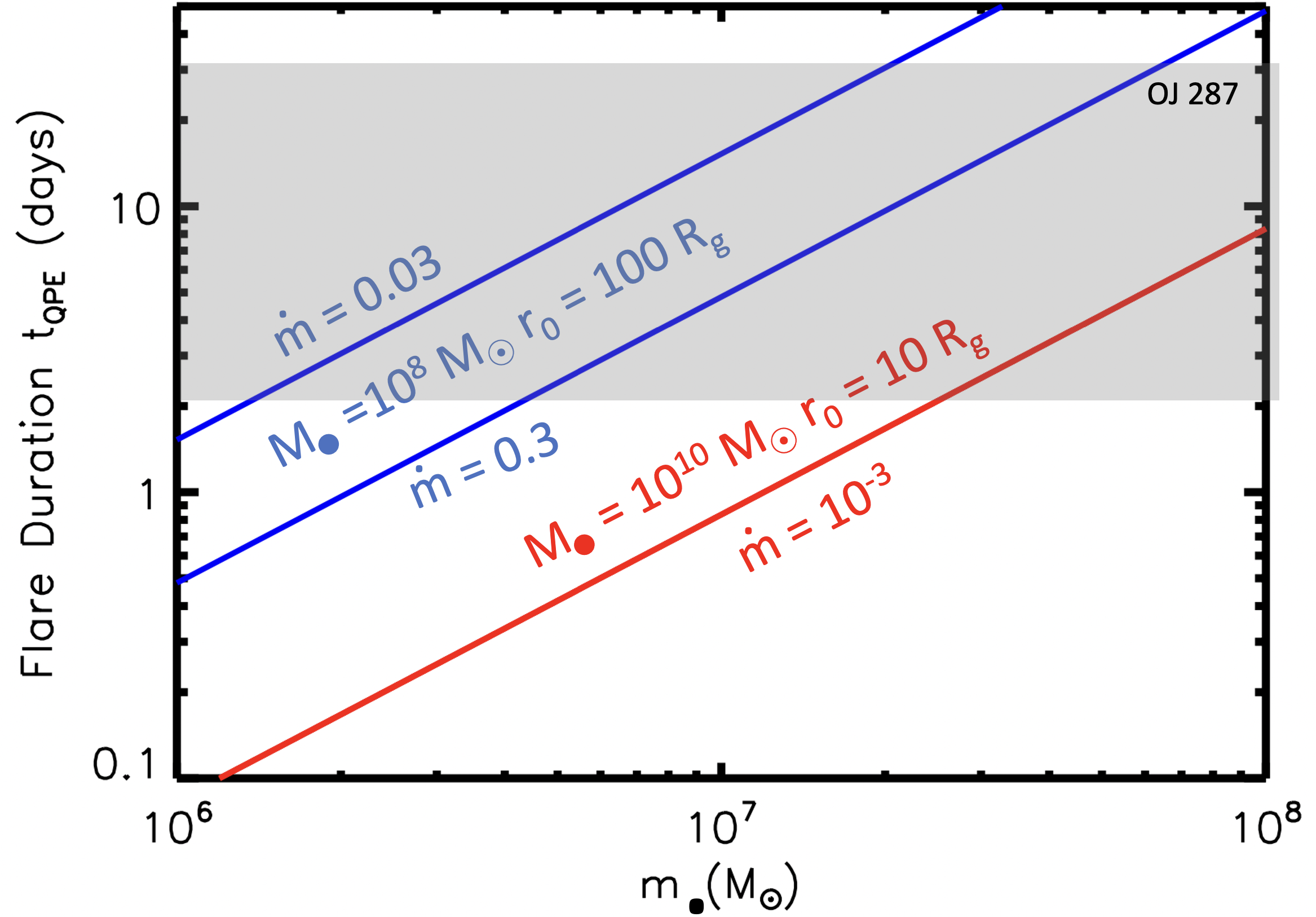}
    \includegraphics[width=0.45\textwidth]{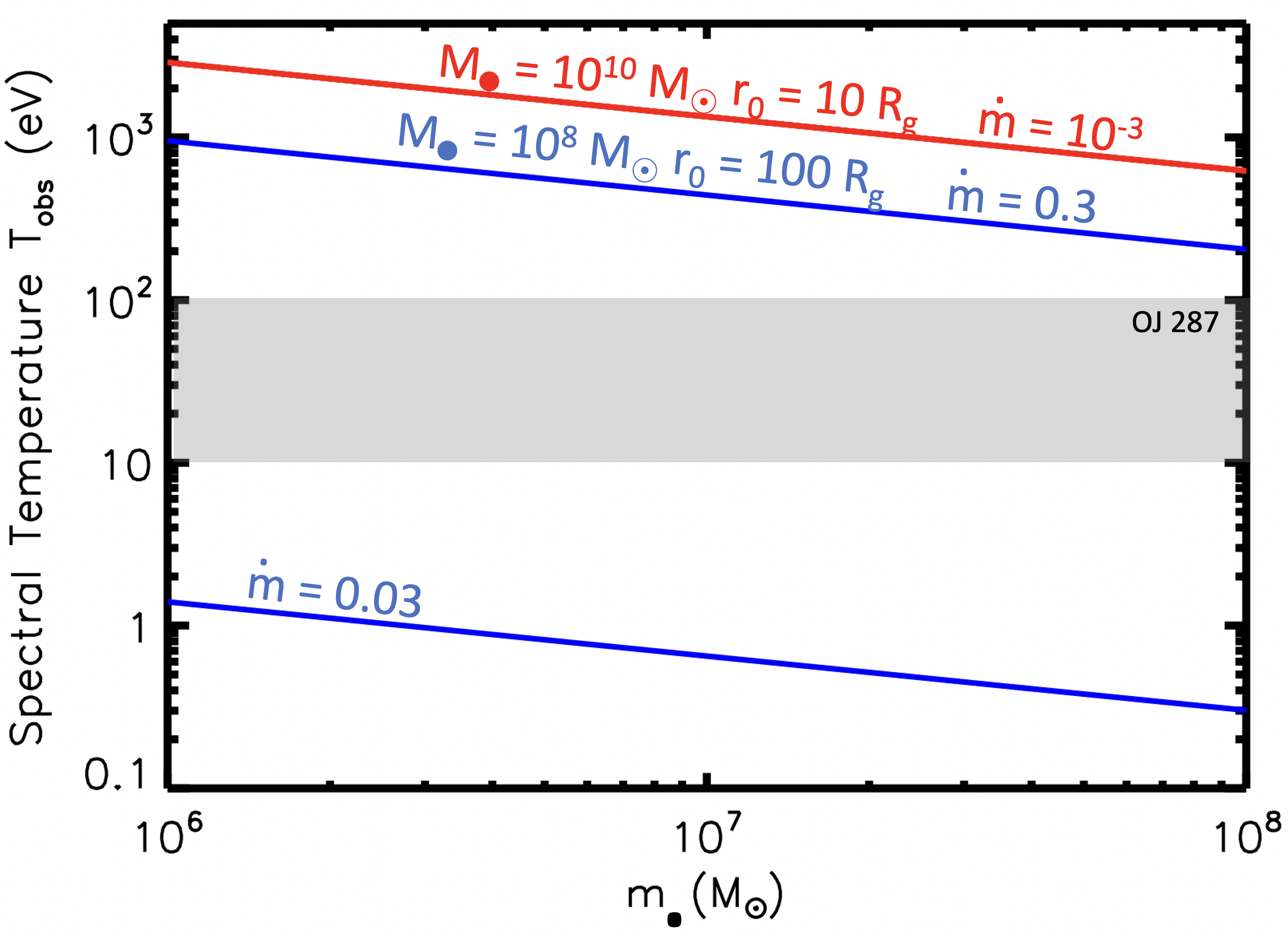}
    \caption{Characteristic flare luminosity (top left; Eq.~\eqref{eq:LpkBH}), duration (top right; Eq.~\eqref{eq:tpkBH}), and spectral temperature (bottom; Eq.~\eqref{eq:Tobs}) from the collision of the secondary SMBH of mass $m_{\bullet}$ with the accretion disk of the primary SMBH of mass $M_{\bullet} = 10^{8}M_{\odot}$ (blue lines) or $M_{\bullet} = 10^{10}M_{\odot}$ (red lines) at a radius $r_0 = 100R_{\rm g}(10R_{\rm g})$; these roughly correspond to the pericenter distance of the hypothesized $\{e \approx 0.7$, $P_{\rm orb} \approx 9$ yr$\}$ orbit of the SMBH binary candidate OJ 287.  Shown for comparison in the gray shaded regions are the range of observed properties of previous optical/UV flares from OJ287 (e.g., \citealt{Valtonen+08,Valtonen+12}).  We assume $\alpha = 0.1$ and $R_{\rm isco} = 4 R_{\rm g}$ for the viscosity and inner radius, respectively, of the quiescent disk.  Although the low-$M_{\bullet}$ models can fit the durations and spectral temperature of the OJ 287 flares, they are generally not sufficiently luminous.}
    \label{fig:LpkTobsBH}
\end{figure}

Taking $M_{\bullet} \sim 10^{8}(10^{10})M_{\odot}$ for the primary SMBH \citep{Dey+18,Komossa+23}, the 9-year rest-frame period implies a semi-major axis $a \sim 10^{3}(10^{2})R_{\rm g},$ but collision radii $R_{\rm p} \simeq a(1-e) \sim 10^{2}(10)R_{\rm g}$, making the OJ 287 system somewhat akin to a scaled-up version of the QPE problem.  Rescaling our predictions for the flare timescale and luminosity (Eqs.~\eqref{eq:tpkBH}, \eqref{eq:LpkBH}) to a much larger secondary mass, we find:
\be
t_{\rm QPE,\bullet} \approx 8\,{\rm d}\,\,\,\frac{(1+z)}{\alpha_{-1}^{1/2}\dot{m}_{-1}^{1/2}}\left(\frac{m_{\bullet}}{10^{7}M_{\odot}}\right)\left(\frac{r_0}{100R_{\rm g}}\right)^{2};
\label{eq:tQPEOJ}
\ee
\begin{eqnarray}
L_{\rm QPE,\bullet} &\approx& 5.5\times 10^{44}\,{\rm erg\,s^{-1}}\left(\frac{m_{\bullet}}{10^{7}M_{\odot}}\right)^{2/3} \times \nonumber \\
&&
\left(\frac{M_{\bullet}}{10^{9}M_{\odot}}\right)^{1/3}\dot{m}_{-1}^{1/3}\left(\frac{r_0}{100R_{\rm g}}\right)^{-1/3},
\label{eq:LQPEOJ}
\end{eqnarray}
where we have neglected cosmological redshift effects on the luminosity.  These flare properties are summarized in Fig.~\ref{fig:LpkTobsBH} along with their predicted spectral temperatures as a function of $m_{\bullet}$ for different primary masses $M_{\bullet} = 10^{8}(10^{10})M_{\odot}$, collision radii $r = 10(100) R_{\rm g}$, and accretion rates $\dot{m} = 0.03-0.3(10^{-3})$.

The major flares from OJ 287 in 1983, 2005, and 2007 exhibited peak durations ranging from days to months \citep{Valtonen+08}, consistent with the predicted $t_{\rm QPE,\bullet} \propto m_{\bullet}r_0^{2}$ (Eq.~\eqref{eq:tQPEOJ}) given the range of secondary masses $m_{\bullet} \sim 10^{6}(10^{8})M_{\odot}$ and corresponding collision $r_0 \sim 10(100) R_{\rm g}$, allowing for some variations in the latter driven by orbital precession effects \citep{Valtonen+08}.  The optical/UV spectrum observed during the $\sim$ month-long 2005 flare is consistent with being optically-thin bremsstrahlung emission of temperature $T_{\rm obs} \sim 10-100$ eV \citep{Valtonen+12}.  While this temperature broadly agrees with the $M_{\bullet} = 10^{8}M_{\odot}$; $r_0 = 100R_{\rm g}$ model (Fig.~\ref{fig:LpkTobsBH}), the observed flare luminosity $\sim 10^{46}$ erg s$^{-1}$ exceeds that predicted (Eq.~\eqref{eq:LQPEOJ}) by over an order of magnitude for the same parameters ($\dot{m} \sim 0.1, M_{\bullet} \sim 10^{8}M_{\odot}, m_{\bullet} \sim 10^{7}M_{\odot})$ that can reproduce the flare duration and temperature.  This conclusion appears to challenge black hole-disk collisions as the source of the optical/UV flares from OJ 287, though a more detailed comparison between the properties of individual flares is warranted.    

The model presented in this paper (Sec.~\ref{sec:lightcurve}; Appendix \ref{sec:BHcollision}) differs in important ways from earlier black hole-disk collision light curve models applied to OJ 287 (e.g., \citealt{Lehto&Valtonen96,Valtonen+12}).  For reasons described in Sec.~\ref{sec:lightcurve}, we assume the shocked debris expands ballistically from the disk surface (ejecta radius $R_{\rm ej} \propto t$), while \citet{Lehto&Valtonen96} limit the expansion rate to the instantaneous sound speed of the adiabatically-cooling gas ($R_{\rm ej} \propto t^{2/3}$).  As in models of supernovae and other explosive transients, we assume that the observed emission peaks once the photon diffusion timescale through the ejecta becomes less than the expansion time ($\tau \sim c/v_{\rm ej}$), while these earlier works assume peak emission occurs only later once $\tau = 1.$  

\section{Discussion and Conclusions}
\label{sec:conclusions}

Building on earlier suggestions \citep{Xian+21,Sukova+21,Miniutti+23}, we have explored periodic collisions between a stellar EMRI and a SMBH accretion disk as a model for QPEs.  Our conclusions are summarized as follows:
\begin{itemize}
    \item  The star's passage through the disk generates a barbell-shaped structure of two hot clouds of shock-heated debris that expand above and below the disk surface, respectively, at close to the orbital speed $v_{\rm K} \sim 0.1c$ (e.g., \citealt{Ivanov+98}).  Although the debris is initially highly optically-thick, radiation begins to escape from the expanding material on a timescale of hours or less.  For orbital radii commensurate with QPE periods, the predicted duration and luminosity of the resulting transient broadly agree with observed QPE flares for a solar-type star as the orbiting body (Fig.~\ref{fig:LpkTobs}).  By contrast, a compact object (white dwarf, neutron star, or stellar-mass black hole) is excluded, insofar that its effective size for interaction with the disk ($\sim$ Bondi radius) is much smaller, resulting in a transient too rapid and dim compared to QPE observations (Appendix \ref{sec:BHcollision}).  Debris from the collision traveling at $v_{\rm K}$ will reach the central regions of the disk on a timescale $\delta t_{\rm Q} \sim P_{\rm QPE}/\pi,$ possibly giving rise to the delayed QPO-like emission observed between the QPE flares in GSN 069 which indeed occur on this timescale (Sec.~\ref{sec:GSN069}).  
    
    \item Since two collisions occur per orbit, and (half of) the debris is visible from both sides of the disk (i.e., for most observer viewing angles), a moderately eccentric orbit naturally gives rise to alternating long-short recurrence interval between QPE flares \citep{Xian+21,Miniutti+23,Franchini+23}.  The star likely migrated to the galactic nucleus through gravitational wave radiation as an EMRI (insofar that the alternative$-$a core left over from a partial TDE$-$would be placed onto a much more eccentric orbit).  Star-disk collisions act to damp away a mild eccentricity over time (Appendix \ref{sec:eccentricity_evolution}), indicating that eccentricities of the QPE-generating stars are likely to be primordial, existing even before the creation of the gaseous disk.  Because the timescale for eccentricity damping can be as short as a decade, alternating long-short QPE systems may evolve over time into QPEs which exhibit a single recurrence time, enabling the possibility of an ``unification scheme'' between the two fledgling QPE behaviors \citep{Arcodia+22}.  General relativistic precession of the orbit is also expected over weeks to years, leading to secular evolution of the flare recurrence time and amplitude (Sec.~\ref{sec:star}).   

\item The disk collision model naturally reproduces the long-short alternating recurrence time observed in GSN-069 \citep{Xian+21}, with the typical $\sim 10\%$ difference between the timing of consecutive flares consistent with the expected residual eccentricity for a stellar EMRI \citep{Linial&Sari22}.  However, the observed correlation between the flare brightness and the subsequent recurrence time (i.e., long interval following a bright flare and vice-versa, \citealt{Miniutti+23}) is not directly addressed within our model. One possible explanation for amplitude differences between consecutive flares is due to asymmetry in the ejecta produced on either side of the disk following star passage.  Alternatively, the orbit's mild eccentricity implies that the star's two collision sites with the disk occur at radii different by order $e$, varying the disk surface density and collision velocity by a similar factor. These interpretations would imply that for a different observer angle (i.e., on the opposite side of the disk) the recurrence interval-flare amplitude correlation would be reversed. Whether this correlation is observer dependent or intrinsic, will be clarified as more QPE sources are detected in the future.
    
    \item Similar to the supernova shock break-out from a compact star (e.g., \citealt{Katz+10,Nakar&Sari10}), photon production in the expanding debris is not rapid enough to maintain thermal equilibrium; as a result, the radiation temperature is considerably harder than the blackbody value, thus enabling the flare radiation to stick out above the softer quiescent disk emission.  Peak spectral energies compatible with those of QPE flares are possible for SMBH masses and accretion rates in the ranges consistent with QPE host galaxies and quiescent disk emission, respectively (Fig.~\ref{fig:LpkTobs}).  However, given the sensitive dependence of the photon production rate and hence peak energy on gas density, $T_{\rm obs} \propto \eta^{2} \propto \rho^{-4}$ (Eq.~\eqref{eq:Tobs}), multidimensional radiation hydrodynamic simulations of star-disk collisions, which explore a potentially more detailed model for the accretion disk and its vertical structure (which remains highly uncertain theoretically; e.g., \citealt{Jiang+19,Mishra+22}), are needed to generate more accurate light curve models and to explore their dependence on viewing angle and the disk-star inclination.  In contrast to the emission temperature, the flare luminosity and duration, depend less sensitively on the details of the disk structure (see Eqs.~\eqref{eq:tpk}, \eqref{eq:Lpk}, and surrounding discussion). 
    
    \item Although the QPE-generating EMRIs likely migrate into the galactic nucleus slowly from larger radii, the strong inverse dependence of the flare temperature on QPE period (Fig.~\ref{fig:LpkTobs}) may bias the observed period distribution to shorter values.  Contributing as well may be the requirement for the stellar orbit to reside close to the circularization radius of the TDE debris, the most likely source of the quiescent gas disk.  Similar to other QPE models which invoke main-sequence EMRIs \citep{Metzger+22,Linial&Sari22,Krolik&Linial22,Lu&Quataert22}, the minimum QPE period is set by the condition for Roche lobe overflow (Eq.~\eqref{eq:TQPEmin}).  
    
    \item The lifetime of QPE activity is generally limited by ram-pressure stripping/ablation of the star to of order a decade to a century (Secs.~\ref{sec:lifetime}, \ref{sec:destruction}).  This may preclude the migration of stellar EMRIs on inclined orbits through long-lived AGN disks, supporting transients like TDEs as the sources of the quiescent gas disk.  This is consistent with the limited radial extent of QPE disks comparable to the TDE circularization radius (e.g., \citealt{Arcodia+21}) and the TDE flares which preceded QPE activity in GSN 069 \citep{Shu+18,Miniutti+19,Miniutti+23} and XMMSL1 J024916.6-041244 \citep{Chakraborty+21}.  Indeed, some QPE sources are associated with young post-starburst galaxies (e.g., \citealt{Wevers+22}), the same kind which preferentially host TDE (e.g., \citealt{Arcavi+14,French+16,Graur+18}), possibly as a result of atypically high nuclear stellar densities \citep{Stone&Metzger16,Stone+18}.       

    \item The relative rates of TDEs and stellar-EMRIs are such that a significant fraction $\sim 1-10\%$ of TDE flares should host a QPE, depending in the details on the EMRI rate and number of TDE flares a given EMRI can survive (Sec.~\ref{sec:rates}).  Within large theoretical and observational uncertainties, the occurrence rate of EMRI-TDE interactions is consistent with the per-galaxy QPE occupation fraction inferred from eROSITA \citep{Arcodia+21}.

    \item Mass stripping/ablation by disk collisions will remove the star's envelope as soon as a few years to a decade following the TDE.  The rate of mass-loss will likely accelerate near the end of this process due to mass-loss-driven radial expansion and potential tidal overflow onto the SMBH, potentially causing the process to evolve towards a singular destruction or envelope-removal event similar to a TDE.  This offers a new explanation for the second TDE-like accretion-flare observed from GSN 069  \citep{Miniutti+23} which naturally explains its timing relative to the first TDE as well as to the cessation of regular QPE activity that occurred nearly simultaneously.  
    
    More generally, sudden TDE-triggered EMRI envelope-removal events offers a mechanism to trigger a second TDE-like outburst in galaxies years to decades after the first outburst, provided an alternative to the popular partial disruption hypothesis for other ``repeating'' TDEs \citep{Campana+15,Liu+23,Malyali+23,Wevers+23}.  
    Even if the EMRI's envelope is removed by the disk in a dynamical event, longer-term survival of its core, and later resumption of QPE activity, cannot be excluded.  
    
    \item Although aspects of our model resemble those which attribute QPE X-ray emission to shocks associated with the circularization of tidally stripped matter from a stellar EMRI on a mildly eccentric orbit \citep{Krolik&Linial22,Lu&Quataert22}, this class of models are in tension with the observed long-short oscillations, which are more easily explained with two flares occurring per orbital period.  The disk-collision model presented here does not require (and in fact does not permit) the star to be overflowing its Roche radius at the time of both flares.

    \item We have mainly focused on stellar orbits with low-eccentricies $e \lesssim 0.1$, motivated by QPE observations and the typical expected residual eccentricity of EMRIs formed through the Hills mechanism.  However, our predictions for the observable properties of disk-collision flares can readily be generalized to stars on more highly eccentric orbits (Sec.~\ref{sec:highe}).  
    \item Although we have focused on the known QPE sources in galactic nuclei hosting low-mass SMBH, our flare emission model can in principle also be applied to stars or compact objects orbiting through the disks of more canonical AGN-hosting $\sim 10^{8}M_{\odot}$ SMBH.  For the same recurrence time and star properties, QPE-like flares from more massive SMBH are predicted to be more luminous, harder, and shorter in duration (Sec.~\ref{sec:lightcurve}).  On the other hand, insofar that TDEs become less frequent or absent altogether from more massive SMBH (e.g., \citealt{Kesden12,vanVelzen18}), the source of the gaseous disk is less clear in this case; as already mentioned, a star's ablation-limited lifetime is much shorter than that of a typical AGN, making it unclear how a low-eccentricity star could make it to short orbital periods in the first place.  This may favor compact object EMRIs such as IMBH as the orbiting bodies responsible for QPE flares in long-lived AGN disks, or alternative mechanisms to generate transient gaseous disks than TDEs (e.g., changing-look AGN).    

    \item Our model can also be applied to the disk collisions in SMBH binaries, such as OJ 287 (Sec.~\ref{sec:OJ287}), giving commensurately longer flare durations and higher flare luminosities (Eqs.~\eqref{eq:tpkBH}, \eqref{eq:LpkBH}).  However, when applied to the optical/UV flares seen from OJ 287 over the past several decades, we find it may be challenging to self-consistently produce all the observed flare properties (Fig.~\ref{fig:LpkTobsBH}).
\end{itemize}

During the final preparation of this manuscript, we became aware of a work by \citet{Tagawa&Haiman23} that explores the radiation signal produced by the shock break-out of a star from an AGN disk, similar in broad terms to the scenario studied in this paper.  Although aspects of our models appear to differ, the essential conclusion of these authors, namely that star-disk collisions can explain observed QPE flares, broadly agree with our own. The latest published observations of GSN-069, \citep{Miniutti+23b} have appeared during the last stages of our paper's review. These new observations reveal the reappearance of QPE flares in July 2022, several months after their disappearance, reported in \citet{Miniutti+23}. In the framework of our model, this observation may indicate the survival of the stellar EMRI despite the second rise in the quiescence. As we have speculated in Sec.~\ref{sec:GSN069}, the re-brightening of the quiescence in late 2021 May occurred due to the partial stripping of mass from the outer layers of the star. The star's surviving core then interacts with the accretion flow, resulting in detectable QPEs flares once the quiescent emission becomes sufficiently faint as the accretion rate drops. It is worth noting that the interval between the two observed flares in the July 2022 epoch is roughly $20 \, \rm ks$, compared to the previous recurrence time of roughly $32 \,\rm ks$ \citep{Miniutti+19,Miniutti+23}. Additional observations of the system will reveal whether the sum of two consecutive recurrence times has changed considerably, or remained similar to the $\approx 64 \, \rm ks$ previously observed. These observations will allow to constrain the evolution of the orbit's semi-major axis and eccentricity, as well as its precession, over the phase during which no QPE flares were observed from this system.

\acknowledgements We are indebted to Riccardo Arcodia, Margherita Giustini, and Giovanni Miniutti for insightful comments and information on QPE sources.  We also acknowledge fruitful discussions with Jordy Davelaar, Zoltan Haiman, Julian Krolik, Yuri Levin, Jerry Ostriker, and Eliot Quataert.  IL acknowledges support from a Rothschild Fellowship and The Gruber Foundation. BDM was supported in part by the National Science Foundation (grant No. AST-2009255). The Flatiron Institute is supported by the Simons Foundation.

\appendix

\section{Flares from Compact Object-Disk Collisions}
\label{sec:BHcollision}

Unlike the case of a non-degenerate stellar EMRI, a compact object such as a stellar-mass black hole of mass $m_{\bullet}$ on a circular orbit of semi-major axis $r_0$, will interact with the gas in the SMBH accretion disk with an effective radius equal to the Bondi radius,
\be
R_{\rm B} \simeq \frac{Gm_{\bullet}}{v_{\rm K}^{2}} \simeq \frac{m_{\bullet}}{M_{\bullet}}r_0 \simeq 0.02R_{\odot}M_{\bullet,6}^{-2/3}\left(\frac{m_{\bullet}}{100M_{\odot}}\right)\mathcal{P}_{\rm QPE,4}^{2/3}.
\ee
Following the same derivation as in Sec.~\ref{sec:lightcurve} but replacing $R_{\star}$ with $R_{\rm B}$ in Eqs.~\eqref{eq:tpk}, \eqref{eq:Lpk}, we see that the timescale and luminosity of the transient emission are given, respectively, by
\be
t_{\rm QPE,\bullet} \approx 2\times 10^{-3}\,{\rm hr}\,\alpha_{-1}^{-1/2}\dot{m}_{-1}^{-1/2}M_{\bullet,6}^{-4/3}\left(\frac{m_{\bullet}}{100M_{\odot}}\right)\mathcal{P}_{\rm QPE,4}^{4/3};
\label{eq:tpkBH}
\ee
\be
L_{\rm QPE,\bullet} \simeq \frac{L_{\rm Edd}}{3}\left(\frac{h}{r_{0}}\right)^{1/3}\left(\frac{m_{\bullet}}{M_{\bullet}}\right)^{2/3} \approx 2.6\times 10^{40}\,{\rm erg\,s^{-1}}\,\dot{m}_{-1}^{1/3}M_{\bullet,6}^{5/9}\left(\frac{m_{\bullet}}{100M_{\odot}}\right)^{2/3}\mathcal{P}_{\rm QPE,4}^{-2/9},
\label{eq:LpkBH}
\ee
both of which are too small to explain observed QPE sources, except for $m_{\bullet} \gtrsim 10^{3}M_{\odot}$ (however, IMBH models for QPE flares can be excluded based on the rate implications; Sec.~\ref{sec:OJ287}).  The spectral temperature of the flare can likewise be written (Eq.~\eqref{eq:Tobs} for $\eta > 1$)
\begin{eqnarray}
k_{\rm B}T_{\rm obs,\bullet} &\approx& 199\,{\rm eV} \alpha_{-1}^{5/2}\dot{m}_{-1}^{11/4}M_{\bullet,6}^{1/3}\left(\frac{m_{\bullet}}{100M_{\odot}}\right)^{-1/3}\left(\frac{r_0}{100R_{\rm g}}\right)^{-41/6} \nonumber
\\
&\approx& 280\,{\rm eV}\,\alpha_{-1}^{5/2}\dot{m}_{-1}^{11/4}M_{\bullet,6}^{44/9}\left(\frac{m_{\bullet}}{100M_{\odot}}\right)^{-1/3}\mathcal{P}_{\rm QPE,4}^{-41/9}.
\label{eq:TobsBH}
\end{eqnarray}

The compact object can in principle impart energy to its environment by accreting material collected passing through the disk.  The maximum accretion rate onto the compact object as it passes through the disk midplane can be estimated from the Bondi rate: 
\be
\dot{M}_{\rm B} \simeq \frac{4\pi G^{2}m_{\bullet}^{2}\rho_{\rm c}}{v_{\rm K}^{3}}.
\ee
The mass accreted per midplane passage of duration $t_{\rm cross} \simeq 2h/v_{\rm K}$ is therefore 
\be
M_{\rm acc} \simeq \dot{M}_{\rm B}t_{\rm cross} =  \frac{4\pi G^{2}m_{\bullet}^{2}\Sigma}{v_{\rm K}^{4}} = 4\pi \Sigma r_0^{2}\left(\frac{m_{\bullet}}{M_{\bullet}}\right)^{2}.
\ee
This is identical within a factor of order unity to Eq.~\eqref{eq:Mej} for the intercepted mass, again replacing the stellar radius $R_{\star}$ with $R_{\rm B}.$   Assuming the accreted material releases energy into the environment (e.g., via disk winds or a jet) with efficiency $\epsilon = 0.1 \epsilon_{-1},$ the resulting total energy release per passage is 
\be
E_{\rm acc} = \epsilon M_{\rm acc}c^{2} \simeq 4\times 10^{43}\,{\rm erg}\,\frac{\epsilon_{\rm -1}}{\alpha_{-1}\dot{m}_{-1}}\left(\frac{m_{\bullet}}{10^{2}M_{\odot}}\right)^{2}\left(\frac{r_0}{100R_{\rm g}}\right)^{7/2},
\ee
where we have used Eq.~\eqref{eq:tauc} for $\Sigma.$  For a stellar-mass compact object ($m_{\bullet} \lesssim 10^{2}M_{\odot}$), $E_{\rm acc}$ is typically several orders of magnitude lower than the energy radiated per QPE flare, $\sim L_{\rm QPE}t_{\rm QPE} \sim 10^{45}-10^{46}$ erg.  Furthermore, given that the implied accretion rates are highly super-Eddington, the accreted matter may not reach the innermost radius of the accretion flow due to outflows (e.g., \citealt{Blandford&Begelman99}), resulting in an effective efficiency $\epsilon \ll 0.1$.  We conclude that accretion power does not change our conclusion that stellar-mass compact objects are likely incapable of explaining QPE flares.  

\section{Eccentricity Evolution due to Star-Disk Collisions} \label{sec:eccentricity_evolution}
Here we consider the effect of star-disk collisions on the orbital eccentricity of a star of mild eccentricity, as implied by the timing of known QPE sources (see for example discussions in \S \ref{sec:star}, \S \ref{sec:highe}).
Consider a star of mass $M_\star$ on an eccentric orbit, with an eccentricity vector $\vec{e}$, intercepting the disk plane twice per orbit. We approximate collisions as an impulsive change in the star's velocity, given by
\begin{equation}
    \vec{v}' = \vec{v}+\delta\vec{v} \,,
\end{equation}
where $\vec{v}$ and $\vec{v}'$ are the velocity vectors immediately before and after the disk passage. Assuming that the drag force experienced by the star is parallel and opposite to its velocity, we have $\delta \vec{v} = -f \vec{v}$, with $f \ll 1$. Momentum conservation suggests $f \approx M_{\rm ej}/M_\star$, where $M_{\rm ej}$ is disk mass intercepted at the collision (as in Eq. \ref{eq:Mej}). Considering the two collisions occurring per orbit, the net change in the eccentricity vector after one full orbit is, to leading order
\begin{equation}
    \delta \vec{e}_{\rm tot} = -2\left[ (f_1+f_2)\vec{e} + (f_1-f_2)\hat{r}_1 \right] \,,
\end{equation}
where $f_1$ and $f_2$ correspond to the impulsive velocity change at the first and second collisions, and $\hat{r}_1$ is the position unit vector at the first collision. Here we used the fact that the two collisions occur at opposite anomalies, such that $\hat{r}_1 = -\hat{r}_2$. We further neglected apsidal precession, which introduces an $\mathcal{O}(R_{\rm g}/r_0)$ correction to the above expression. The disk conditions at the two collision sites vary by order $\mathcal{O}(|\vec{e}|)$, thus we can write $f_1-f_2 = A \bar{f} e$, where $\bar{f} = (f_1+f_2)/2$, and $A$ is an order unity factor that depends on the relative orientation of the orbit with respect to the disk plane. We thus write
\begin{equation}
    \delta \vec{e}_{\rm tot} = -2\bar{f} \left[ 2 \vec{e} + Ae\hat{r}_1 \right] \,.
\end{equation}
The per-orbit fractional change in the magnitude of the eccentricity vector is, to lowest order in $f$
\begin{equation}
    \frac{\delta e}{e} = \left( \sqrt{ 1 + 2 \frac{\vec{e} \cdot \delta \vec{e}_{\rm tot}}{e^2} } - 1 \right) \approx \frac{\vec{e} \cdot \delta \vec{e}_{\rm tot}}{e^2} = -2\bar{f}\left( 2 + A \cos{\varpi} \right) \,,
\end{equation}
where $\varpi$ is the angle between between the radius vector of the first collision site and the eccentricity vector.
In the above derivation we assumed $f_1,f_2\ll e \lesssim 0.1$, such that higher order terms were omitted. 

It is apparent that the sign of $\delta e/e$ depends on the sign of $(2+A \cos{\varpi})$. For a radiation pressure dominated disk, the disk surface density scales as $\Sigma \propto r^{3/2}$ (e.g., Eq. \ref{eq:tauc}). Since $f \propto \Sigma$, we have $A \lesssim 3/2$, under the assumption that $e\lesssim 0.1$. We conclude that for eccentricities comparable to those required to explain QPE sources, the orbital eccentricity will dampen over time as a result of star-disk collisions. 

During its GW circularization phase, the EMRI's eccentricity gradually evolves from $1-e \ll 1$, to mild value of $e \lesssim 0.5$. As discussed in \S \ref{sec:rates}, multiple TDEs may occur during this high eccentricity phase. However, insofar that the TDE disk is compact, EMRI collisions will only occur around pericenter, decreasing the apocenter distance, yet maintaining a nearly fixed pericenter distance, thereby not contributing to changes in the EMRI's angular momentum. Orbital eccentricity is therefore dampened both in the mild and high $e$ regimes.

\section{Disk Corona from Collision Ejecta}
\label{sec:corona}

The unperturbed disk midplane is not the only gas with which the star will interact each orbit.  The mass ejected by each star-disk collision (Eq.~\eqref{eq:Eej}) possesses a characteristic spread in its specific internal energy $\sim v_{\rm K}^{2}$ comparable to the escape speed and hence will be deposited over an annulus centered around the collision radius $r_{0}$ with a characteristic radial thickness $\sim r_{0}$ and surface area $\sim \pi r_{0}^{2}$.  The accumulation of this hot shocked gas may in some cases form a hot (``coronal'') layer on top of the disk of characteristic vertical scale-height $h_{\rm cor} > h$, sound speed $c_{\rm s,cor} = h_{\rm cor}\Omega$ and surface density $\Sigma_{\rm cor}$.  Here we estimate the properties of this layer, before speculating about its effect on the star-disk collision emission.

The corona is fed mass by star-disk collisions at a per-side rate
\be
\dot{M}_{+} = \frac{M_{\rm ej}}{2P_{\rm QPE}} \simeq \frac{\pi R_{\star}^{2} \Sigma}{P_{\rm QPE}} = \frac{R_{\star}^{2}\Sigma}{r_{0}}v_{\rm K},
\label{eq:Sigmaplus}
\ee 
leading to heating of the corona at a rate
\be
\dot{E}_{+} \simeq \frac{1}{2}\dot{M}_{+}v_{\rm sh}^{2} \simeq  \frac{R_{\star}^{2}\Sigma}{r_{0}}v_{\rm K}^{3}
\label{eq:Edotplus}
\ee 
The corona is radiation-dominated and therefore cools at the Eddington luminosity (corresponding to just the vertical component of gravity),
\be
\dot{E}_{-} \simeq \pi r_{0}^{2}\sigma T_{\rm eff}^{4} \simeq \frac{3}{16}\frac{h_{\rm cor}}{r_{0}}L_{\rm Edd},
\label{eq:Edotminus}
\ee
where we have used $c_{\rm s,cor} \simeq aT_{\rm cor}^{4}/3 \rho_{\rm cor}$, $\rho_{\rm cor} \simeq \Sigma_{\rm cor}/h_{\rm cor}$, $(3/4)T_{\rm cor}^{4} \simeq \tau_{\rm cor} T_{\rm eff}^{4}$, $\tau_{\rm cor} = \Sigma_{\rm cor}\kappa_{\rm T}$.  Equating heating \eqref{eq:Edotplus} and cooling \eqref{eq:Edotminus}, thus gives the corona thickness 
\be
\frac{h_{\rm cor}}{r_{0}} \simeq \frac{16}{3}\frac{R_{\star}^{2}\Sigma v_{\rm K}^{3}}{r_{0}L_{\rm Edd}} \simeq 0.004\frac{\mathcal{R}_{\star}^{2}}{\alpha_{-1}\dot{m}_{-1}M_{\bullet,6}^{2}}\left(\frac{r_{0}}{100 R_{\rm g}}\right)^{-1},
\label{eq:hcor1}
\ee
where we have used Eq.~\eqref{eq:tauc}.  Comparing to the midplane scale-height $h$ (Eq.~\eqref{eq:hoverr}), 
\be
\frac{h_{\rm cor}}{h} \simeq 0.28\frac{\mathcal{R}_{\star}^{2}}{\alpha_{-1}\dot{m}_{-1}^{2}M_{\bullet,6}^{2}}
\label{eq:hcor}
\ee
we see that the corona will significantly expand the disk thickness relative to the standard case, for a low accretion rates, large star, and/or a low-mass SMBH.  

As with the midplane, the surface density of the coronal layer is set by the rate the deposited matter accretes inwards due to viscosity:
\be
\dot{M}_- \simeq 3\pi \nu_{\rm cor} \Sigma_{\rm cor},
\ee
where now $\nu_{\rm cor} = \alpha (h_{\rm cor}/r_{0})^{2}(GM_{\bullet} r_{0})^{1/2}$ is the viscosity of the coronal layer.  Equating $\dot{M}_+ = \dot{M}_-$ gives the surface density of the corona 
\begin{eqnarray}
\frac{\Sigma_{\rm cor}}{\Sigma} \simeq \frac{3}{8\pi \alpha}\frac{L_{\rm Edd}^{2}}{R_{\star}^{2}\Sigma^{2}v_{\rm K}^{6}} \approx 7\alpha_{-1}\dot{m}_{-1}^{2}\mathcal{R}_{\star}^{-2}M_{\bullet,6}^{2}
\label{eq:Sigmacor}
\end{eqnarray}

To estimate what effect the corona will have on the star-disk collision emission (Sec.~\ref{sec:lightcurve}), consider that the flare timescale and luminosity obey (Eqs.~\eqref{eq:tpk}, \eqref{eq:Lpk})
\be
P_{\rm QPE} \propto \Sigma^{1/2};\,\,\,\,
L_{\rm QPE} \propto h^{1/3}.
\ee
Eqs.~\eqref{eq:hcor}, \eqref{eq:Sigmacor} show that when $h_{\rm cor} \gtrsim h$ is satisfied, we typically have $\Sigma_{\rm cor} \lesssim \Sigma$.  Thus, although the presence of a distinct coronal layer ($h_{\rm cor} > h$) will sharpen the light curve peak modestly by increasing $L_{\rm QPE}$ and reducing $P_{\rm QPE}$ (for emission originating from the shocked uppermost layers), it is unlikely to substantially alter the total radiated energy or duration of the flare, which are still dominated by the shocked midplane material.


\begin{thebibliography}{}
\expandafter\ifx\csname natexlab\endcsname\relax\def\natexlab#1{#1}\fi
\providecommand{\url}[1]{\href{#1}{#1}}
\providecommand{\dodoi}[1]{doi:~\href{http://doi.org/#1}{\nolinkurl{#1}}}
\providecommand{\doeprint}[1]{\href{http://ascl.net/#1}{\nolinkurl{http://ascl.net/#1}}}
\providecommand{\doarXiv}[1]{\href{https://arxiv.org/abs/#1}{\nolinkurl{https://arxiv.org/abs/#1}}}

\bibitem[{{Arcavi} {et~al.}(2014)}]{Arcavi+14}
{Arcavi}, I., {et~al.} 2014, \apj, 793, 38, \dodoi{10.1088/0004-637X/793/1/38}

\bibitem[{{Arcodia} {et~al.}(2021){Arcodia}, {Merloni}, {Nandra}, {Buchner},
  {Salvato}, {Pasham}, {Remillard}, {Comparat}, {Lamer}, {Ponti}, {Malyali},
  {Wolf}, {Arzoumanian}, {Bogensberger}, {Buckley}, {Gendreau}, {Gromadzki},
  {Kara}, {Krumpe}, {Markwardt}, {Ramos-Ceja}, {Rau}, {Schramm}, \&
  {Schwope}}]{Arcodia+21}
{Arcodia}, R., {Merloni}, A., {Nandra}, K., {et~al.} 2021, \nat, 592, 704,
  \dodoi{10.1038/s41586-021-03394-6}

\bibitem[{{Arcodia} {et~al.}(2022){Arcodia}, {Miniutti}, {Ponti}, {Buchner},
  {Giustini}, {Merloni}, {Nandra}, {Vincentelli}, {Kara}, {Salvato}, \&
  {Pasham}}]{Arcodia+22}
{Arcodia}, R., {Miniutti}, G., {Ponti}, G., {et~al.} 2022, \aap, 662, A49,
  \dodoi{10.1051/0004-6361/202243259}

\bibitem[{{Armitage} {et~al.}(1996){Armitage}, {Zurek}, \&
  {Davies}}]{Armitage+96}
{Armitage}, P.~J., {Zurek}, W.~H., \& {Davies}, M.~B. 1996, \apj, 470, 237,
  \dodoi{10.1086/177864}

\bibitem[{{Arnett}(1980)}]{Arnett80}
{Arnett}, W.~D. 1980, \apj, 237, 541, \dodoi{10.1086/157898}

\bibitem[{{Auchettl} {et~al.}(2017){Auchettl}, {Guillochon}, \&
  {Ramirez-Ruiz}}]{Auchettl+17}
{Auchettl}, K., {Guillochon}, J., \& {Ramirez-Ruiz}, E. 2017, \apj, 838, 149,
  \dodoi{10.3847/1538-4357/aa633b}

\bibitem[{{Blaes} {et~al.}(2006){Blaes}, {Davis}, {Hirose}, {Krolik}, \&
  {Stone}}]{Blaes+06}
{Blaes}, O.~M., {Davis}, S.~W., {Hirose}, S., {Krolik}, J.~H., \& {Stone},
  J.~M. 2006, \apj, 645, 1402, \dodoi{10.1086/503741}

\bibitem[{{Blandford} \& {Begelman}(1999)}]{Blandford&Begelman99}
{Blandford}, R.~D., \& {Begelman}, M.~C. 1999, \mnras, 303, L1,
  \dodoi{10.1046/j.1365-8711.1999.02358.x}

\bibitem[{{Bonnerot} \& {Lu}(2020)}]{Bonnerot&Lu20}
{Bonnerot}, C., \& {Lu}, W. 2020, \mnras, 495, 1374,
  \dodoi{10.1093/mnras/staa1246}

\bibitem[{{Bortolas} {et~al.}(2023){Bortolas}, {Ryu}, {Broggi}, \&
  {Sesana}}]{Bortolas+23}
{Bortolas}, E., {Ryu}, T., {Broggi}, L., \& {Sesana}, A. 2023, arXiv e-prints,
  arXiv:2303.03408, \dodoi{10.48550/arXiv.2303.03408}

\bibitem[{{Campana} {et~al.}(2015){Campana}, {Mainetti}, {Colpi}, {Lodato},
  {D'Avanzo}, {Evans}, \& {Moretti}}]{Campana+15}
{Campana}, S., {Mainetti}, D., {Colpi}, M., {et~al.} 2015, \aap, 581, A17,
  \dodoi{10.1051/0004-6361/201525965}

\bibitem[{{Cannizzo} {et~al.}(1990){Cannizzo}, {Lee}, \&
  {Goodman}}]{Cannizzo+90}
{Cannizzo}, J.~K., {Lee}, H.~M., \& {Goodman}, J. 1990, \apj, 351, 38,
  \dodoi{10.1086/168442}

\bibitem[{{Chakraborty} {et~al.}(2021){Chakraborty}, {Kara}, {Masterson},
  {Giustini}, {Miniutti}, \& {Saxton}}]{Chakraborty+21}
{Chakraborty}, J., {Kara}, E., {Masterson}, M., {et~al.} 2021, \apjl, 921, L40,
  \dodoi{10.3847/2041-8213/ac313b}

\bibitem[{{Chen} {et~al.}(2022){Chen}, {Shen}, \& {Liu}}]{Chen+22}
{Chen}, J.-H., {Shen}, R.-F., \& {Liu}, S.-F. 2022, arXiv e-prints,
  arXiv:2210.09945, \dodoi{10.48550/arXiv.2210.09945}

\bibitem[{{Cufari} {et~al.}(2023){Cufari}, {Nixon}, \& {Coughlin}}]{Cufari+23}
{Cufari}, M., {Nixon}, C.~J., \& {Coughlin}, E.~R. 2023, \mnras, 520, L38,
  \dodoi{10.1093/mnrasl/slad001}

\bibitem[{{Dai} {et~al.}(2010){Dai}, {Fuerst}, \& {Blandford}}]{Dai+10}
{Dai}, L.~J., {Fuerst}, S.~V., \& {Blandford}, R. 2010, \mnras, 402, 1614,
  \dodoi{10.1111/j.1365-2966.2009.16038.x}

\bibitem[{{Dey} {et~al.}(2018){Dey}, {Valtonen}, {Gopakumar}, {Zola},
  {et~al.}}]{Dey+18}
{Dey}, L., {Valtonen}, M.~J., {Gopakumar}, A., {Zola}, S., {et~al.} 2018, \apj,
  866, 11, \dodoi{10.3847/1538-4357/aadd95}

\bibitem[{{Franchini} {et~al.}(2023){Franchini}, {Bonetti}, {Lupi}, {Miniutti},
  {Bortolas}, {Giustini}, {Dotti}, {Sesana}, {Arcodia}, \&
  {Ryu}}]{Franchini+23}
{Franchini}, A., {Bonetti}, M., {Lupi}, A., {et~al.} 2023, arXiv e-prints,
  arXiv:2304.00775, \dodoi{10.48550/arXiv.2304.00775}

\bibitem[{{Frank} {et~al.}(2002){Frank}, {King}, \& {Raine}}]{Frank+02}
{Frank}, J., {King}, A., \& {Raine}, D.~J. 2002, {Accretion Power in
  Astrophysics: Third Edition}

\bibitem[{{French} {et~al.}(2016){French}, {Arcavi}, \&
  {Zabludoff}}]{French+16}
{French}, K.~D., {Arcavi}, I., \& {Zabludoff}, A. 2016, \apjl, 818, L21,
  \dodoi{10.3847/2041-8205/818/1/L21}

\bibitem[{{Gafton} {et~al.}(2015){Gafton}, {Tejeda}, {Guillochon}, {Korobkin},
  \& {Rosswog}}]{Gafton+15}
{Gafton}, E., {Tejeda}, E., {Guillochon}, J., {Korobkin}, O., \& {Rosswog}, S.
  2015, \mnras, 449, 771, \dodoi{10.1093/mnras/stv350}

\bibitem[{{Giustini} {et~al.}(2020){Giustini}, {Miniutti}, \&
  {Saxton}}]{Giustini+20}
{Giustini}, M., {Miniutti}, G., \& {Saxton}, R.~D. 2020, \aap, 636, L2,
  \dodoi{10.1051/0004-6361/202037610}

\bibitem[{{Graur} {et~al.}(2018){Graur}, {French}, {Zahid}, {Guillochon},
  {Mandel}, {Auchettl}, \& {Zabludoff}}]{Graur+18}
{Graur}, O., {French}, K.~D., {Zahid}, H.~J., {et~al.} 2018, \apj, 853, 39,
  \dodoi{10.3847/1538-4357/aaa3fd}

\bibitem[{{Guillochon} \& {Ramirez-Ruiz}(2013)}]{Guillochon&RamirezRuiz13}
{Guillochon}, J., \& {Ramirez-Ruiz}, E. 2013, \apj, 767, 25,
  \dodoi{10.1088/0004-637X/767/1/25}

\bibitem[{{Guillochon} \& {Ramirez-Ruiz}(2015)}]{Guillochon&RamirezRuiz15}
---. 2015, ArXiv e-prints.
\newblock \doarXiv{1501.05306}

\bibitem[{{Hills}(1975)}]{Hills75}
{Hills}, J.~G. 1975, \nat, 254, 295, \dodoi{10.1038/254295a0}

\bibitem[{{Hills}(1988)}]{Hills88}
---. 1988, \nat, 331, 687, \dodoi{10.1038/331687a0}

\bibitem[{{Hirose} {et~al.}(2009){Hirose}, {Blaes}, \& {Krolik}}]{Hirose+09}
{Hirose}, S., {Blaes}, O., \& {Krolik}, J.~H. 2009, \apj, 704, 781,
  \dodoi{10.1088/0004-637X/704/1/781}

\bibitem[{{Illarionov} \& {Siuniaev}(1975)}]{Illarionov&Siuniaev75}
{Illarionov}, A.~F., \& {Siuniaev}, R.~A. 1975, \sovast, 18, 413

\bibitem[{{Ingram} {et~al.}(2021){Ingram}, {Motta}, {Aigrain}, \&
  {Karastergiou}}]{Ingram+21}
{Ingram}, A., {Motta}, S.~E., {Aigrain}, S., \& {Karastergiou}, A. 2021,
  \mnras, 503, 1703, \dodoi{10.1093/mnras/stab609}

\bibitem[{{Ivanov} {et~al.}(1998){Ivanov}, {Igumenshchev}, \&
  {Novikov}}]{Ivanov+98}
{Ivanov}, P.~B., {Igumenshchev}, I.~V., \& {Novikov}, I.~D. 1998, \apj, 507,
  131, \dodoi{10.1086/306324}

\bibitem[{{Jiang} {et~al.}(2019){Jiang}, {Blaes}, {Stone}, \&
  {Davis}}]{Jiang+19}
{Jiang}, Y.-F., {Blaes}, O., {Stone}, J.~M., \& {Davis}, S.~W. 2019, \apj, 885,
  144, \dodoi{10.3847/1538-4357/ab4a00}

\bibitem[{{Jiang} {et~al.}(2013){Jiang}, {Stone}, \& {Davis}}]{Jiang+13}
{Jiang}, Y.-F., {Stone}, J.~M., \& {Davis}, S.~W. 2013, \apj, 778, 65,
  \dodoi{10.1088/0004-637X/778/1/65}

\bibitem[{{Jonker} {et~al.}(2020){Jonker}, {Stone}, {Generozov}, {van Velzen},
  \& {Metzger}}]{Jonker+20}
{Jonker}, P.~G., {Stone}, N.~C., {Generozov}, A., {van Velzen}, S., \&
  {Metzger}, B. 2020, \apj, 889, 166, \dodoi{10.3847/1538-4357/ab659c}

\bibitem[{{Kasen} {et~al.}(2007){Kasen}, {Woosley}, {Nugent}, \&
  {R{\"o}pke}}]{Kasen+07}
{Kasen}, D., {Woosley}, S., {Nugent}, P., \& {R{\"o}pke}, F. 2007, in Journal
  of Physics Conference Series, Vol.~78, Journal of Physics Conference Series,
  012037

\bibitem[{{Katz} {et~al.}(2010){Katz}, {Budnik}, \& {Waxman}}]{Katz+10}
{Katz}, B., {Budnik}, R., \& {Waxman}, E. 2010, \apj, 716, 781,
  \dodoi{10.1088/0004-637X/716/1/781}

\bibitem[{{Kaur} {et~al.}(2022){Kaur}, {Stone}, \& {Gilbaum}}]{Kaur+22}
{Kaur}, K., {Stone}, N.~C., \& {Gilbaum}, S. 2022, arXiv e-prints,
  arXiv:2211.00704, \dodoi{10.48550/arXiv.2211.00704}

\bibitem[{{Kesden}(2012)}]{Kesden12}
{Kesden}, M. 2012, \prd, 85, 024037, \dodoi{10.1103/PhysRevD.85.024037}

\bibitem[{{King}(2020)}]{King20}
{King}, A. 2020, \mnras, 493, L120, \dodoi{10.1093/mnrasl/slaa020}

\bibitem[{{King}(2022)}]{King22}
---. 2022, \mnras, 515, 4344, \dodoi{10.1093/mnras/stac1641}

\bibitem[{{King}(2023)}]{King23}
---. 2023, \mnras, 520, L63, \dodoi{10.1093/mnrasl/slad006}

\bibitem[{{Kippenhahn} \& {Weigert}(1990)}]{Kippenhahn&Weigert90}
{Kippenhahn}, R., \& {Weigert}, A. 1990, {Stellar Structure and Evolution}

\bibitem[{{Komossa} {et~al.}(2023){Komossa}, {Grupe}, {Kraus}, {Gurwell},
  {Haiman}, {Liu}, {Tchekhovskoy}, {Gallo}, {Berton}, {Blandford}, {G{\'o}mez},
  \& {Gonzalez}}]{Komossa+23}
{Komossa}, S., {Grupe}, D., {Kraus}, A., {et~al.} 2023, \mnras,
  \dodoi{10.1093/mnrasl/slad016}

\bibitem[{{Krolik} \& {Linial}(2022)}]{Krolik&Linial22}
{Krolik}, J.~H., \& {Linial}, I. 2022, \apj, 941, 24,
  \dodoi{10.3847/1538-4357/ac9eb6}

\bibitem[{{Lehto} \& {Valtonen}(1996)}]{Lehto&Valtonen96}
{Lehto}, H.~J., \& {Valtonen}, M.~J. 1996, \apj, 460, 207,
  \dodoi{10.1086/176962}

\bibitem[{{Lightman} \& {Eardley}(1974)}]{Lightman&Eardley74}
{Lightman}, A.~P., \& {Eardley}, D.~M. 1974, \apjl, 187, L1,
  \dodoi{10.1086/181377}

\bibitem[{{Linial} \& {Sari}(2017)}]{Linial&Sari17}
{Linial}, I., \& {Sari}, R. 2017, \mnras, 469, 2441,
  \dodoi{10.1093/mnras/stx1041}

\bibitem[{{Linial} \& {Sari}(2022)}]{Linial&Sari22}
---. 2022, arXiv e-prints, arXiv:2211.09851, \dodoi{10.48550/arXiv.2211.09851}

\bibitem[{{Liu} {et~al.}(2023){Liu}, {Malyali}, {Krumpe}, {Homan}, {Goodwin},
  {Grotova}, {Kawka}, {Rau}, {Merloni}, {Anderson}, {Miller-Jones},
  {Markowitz}, {Ciroi}, {Di Mille}, {Schramm}, {Tang}, {Buckley}, {Gromadzki},
  {Jin}, \& {Buchner}}]{Liu+23}
{Liu}, Z., {Malyali}, A., {Krumpe}, M., {et~al.} 2023, \aap, 669, A75,
  \dodoi{10.1051/0004-6361/202244805}

\bibitem[{{Liu} {et~al.}(2015){Liu}, {Tauris}, {R{\"o}pke}, {Moriya},
  {Kruckow}, {Stancliffe}, \& {Izzard}}]{Liu+15}
{Liu}, Z.-W., {Tauris}, T.~M., {R{\"o}pke}, F.~K., {et~al.} 2015, \aap, 584,
  A11, \dodoi{10.1051/0004-6361/201526757}

\bibitem[{{Lu} \& {Quataert}(2022)}]{Lu&Quataert22}
{Lu}, W., \& {Quataert}, E. 2022, arXiv e-prints, arXiv:2210.08023,
  \dodoi{10.48550/arXiv.2210.08023}

\bibitem[{{MacLeod} \& {Lin}(2020)}]{MacLeod&Lin20}
{MacLeod}, M., \& {Lin}, D. N.~C. 2020, \apj, 889, 94,
  \dodoi{10.3847/1538-4357/ab64db}

\bibitem[{{Magorrian} \& {Tremaine}(1999)}]{Magorrian&Tremaine99}
{Magorrian}, J., \& {Tremaine}, S. 1999, \mnras, 309, 447,
  \dodoi{10.1046/j.1365-8711.1999.02853.x}

\bibitem[{{Malyali} {et~al.}(2023){Malyali}, {Liu}, {Rau}, {Grotova},
  {Merloni}, {Goodwin}, {Anderson}, {Miller-Jones}, {Kawka}, {Arcodia},
  {Buchner}, {Nandra}, {Homan}, \& {Krumpe}}]{Malyali+23}
{Malyali}, A., {Liu}, Z., {Rau}, A., {et~al.} 2023, \mnras, 520, 3549,
  \dodoi{10.1093/mnras/stad022}

\bibitem[{{Manukian} {et~al.}(2013){Manukian}, {Guillochon}, {Ramirez-Ruiz}, \&
  {O'Leary}}]{Manukian+13}
{Manukian}, H., {Guillochon}, J., {Ramirez-Ruiz}, E., \& {O'Leary}, R.~M. 2013,
  \apjl, 771, L28, \dodoi{10.1088/2041-8205/771/2/L28}

\bibitem[{{Merritt}(2010)}]{Merritt10}
{Merritt}, D. 2010, \apj, 718, 739, \dodoi{10.1088/0004-637X/718/2/739}

\bibitem[{{Metzger}(2022)}]{Metzger22b}
{Metzger}, B.~D. 2022, \apjl, 937, L12, \dodoi{10.3847/2041-8213/ac90ba}

\bibitem[{{Metzger} {et~al.}(2022){Metzger}, {Stone}, \&
  {Gilbaum}}]{Metzger+22}
{Metzger}, B.~D., {Stone}, N.~C., \& {Gilbaum}, S. 2022, \apj, 926, 101,
  \dodoi{10.3847/1538-4357/ac3ee1}

\bibitem[{{Miniutti} {et~al.}(2023{\natexlab{a}}){Miniutti}, {Giustini},
  {Arcodia}, {Saxton}, {Chakraborty}, {Read}, \& {Kara}}]{Miniutti+23b}
{Miniutti}, G., {Giustini}, M., {Arcodia}, R., {et~al.} 2023{\natexlab{a}},
  arXiv e-prints, arXiv:2305.09717, \dodoi{10.48550/arXiv.2305.09717}

\bibitem[{{Miniutti} {et~al.}(2023{\natexlab{b}}){Miniutti}, {Giustini},
  {Arcodia}, {Saxton}, {Read}, {Bianchi}, \& {Alexander}}]{Miniutti+23}
---. 2023{\natexlab{b}}, \aap, 670, A93, \dodoi{10.1051/0004-6361/202244512}

\bibitem[{{Miniutti} {et~al.}(2013){Miniutti}, {Saxton},
  {Rodr{\'\i}guez-Pascual}, {Read}, {Esquej}, {Colless}, {Dobbie}, \&
  {Spolaor}}]{Miniutti+13}
{Miniutti}, G., {Saxton}, R.~D., {Rodr{\'\i}guez-Pascual}, P.~M., {et~al.}
  2013, \mnras, 433, 1764, \dodoi{10.1093/mnras/stt850}

\bibitem[{{Miniutti} {et~al.}(2019){Miniutti}, {Saxton}, {Giustini},
  {Alexander}, {Fender}, {Heywood}, {Monageng}, {Coriat}, {Tzioumis}, {Read},
  {Knigge}, {Gandhi}, {Pretorius}, \& {Ag{\'\i}s-Gonz{\'a}lez}}]{Miniutti+19}
{Miniutti}, G., {Saxton}, R.~D., {Giustini}, M., {et~al.} 2019, \nat, 573, 381,
  \dodoi{10.1038/s41586-019-1556-x}

\bibitem[{{Mishra} {et~al.}(2022){Mishra}, {Fragile}, {Anderson},
  {Blankenship}, {Li}, \& {Nalewajko}}]{Mishra+22}
{Mishra}, B., {Fragile}, P.~C., {Anderson}, J., {et~al.} 2022, \apj, 939, 31,
  \dodoi{10.3847/1538-4357/ac938b}

\bibitem[{{Nakar} \& {Sari}(2010)}]{Nakar&Sari10}
{Nakar}, E., \& {Sari}, R. 2010, \apj, 725, 904,
  \dodoi{10.1088/0004-637X/725/1/904}

\bibitem[{{Nayakshin} {et~al.}(2004){Nayakshin}, {Cuadra}, \&
  {Sunyaev}}]{Nayakshin+04}
{Nayakshin}, S., {Cuadra}, J., \& {Sunyaev}, R. 2004, \aap, 413, 173,
  \dodoi{10.1051/0004-6361:20031537}

\bibitem[{{Nixon} \& {Coughlin}(2022)}]{Nixon&Coughlin22}
{Nixon}, C.~J., \& {Coughlin}, E.~R. 2022, \apjl, 927, L25,
  \dodoi{10.3847/2041-8213/ac5118}

\bibitem[{{Pan} {et~al.}(2022){Pan}, {Li}, {Cao}, {Miniutti}, \& {Gu}}]{Pan+22}
{Pan}, X., {Li}, S.-L., {Cao}, X., {Miniutti}, G., \& {Gu}, M. 2022, \apjl,
  928, L18, \dodoi{10.3847/2041-8213/ac5faf}

\bibitem[{{Pihajoki}(2016)}]{Pihajoki+16}
{Pihajoki}, P. 2016, \mnras, 457, 1145, \dodoi{10.1093/mnras/stv3023}

\bibitem[{{Raj} \& {Nixon}(2021)}]{Raj&Nixon21}
{Raj}, A., \& {Nixon}, C.~J. 2021, \apj, 909, 82,
  \dodoi{10.3847/1538-4357/abdc25}

\bibitem[{{Rees}(1988)}]{Rees88}
{Rees}, M.~J. 1988, \nat, 333, 523, \dodoi{10.1038/333523a0}

\bibitem[{{Saxton} {et~al.}(2011){Saxton}, {Read}, {Esquej}, {Miniutti}, \&
  {Alvarez}}]{Saxton+11}
{Saxton}, R., {Read}, A., {Esquej}, P., {Miniutti}, G., \& {Alvarez}, E. 2011,
  arXiv e-prints, arXiv:1106.3507.
\newblock \doarXiv{1106.3507}

\bibitem[{{Semer{\'a}k} {et~al.}(1999){Semer{\'a}k}, {Karas}, \& {de
  Felice}}]{Semerak+99}
{Semer{\'a}k}, O., {Karas}, V., \& {de Felice}, F. 1999, \pasj, 51, 571,
  \dodoi{10.1093/pasj/51.5.571}

\bibitem[{{Shakura} \& {Sunyaev}(1973)}]{Shakura&Sunyaev73}
{Shakura}, N.~I., \& {Sunyaev}, R.~A. 1973, \aap, 24, 337

\bibitem[{{Shen} \& {Matzner}(2014)}]{Shen&Matzner14}
{Shen}, R.-F., \& {Matzner}, C.~D. 2014, \apj, 784, 87,
  \dodoi{10.1088/0004-637X/784/2/87}

\bibitem[{{Sheng} {et~al.}(2021){Sheng}, {Wang}, {Ferland}, {Shu}, {Yang},
  {Jiang}, \& {Chen}}]{Sheng+21}
{Sheng}, Z., {Wang}, T., {Ferland}, G., {et~al.} 2021, \apjl, 920, L25,
  \dodoi{10.3847/2041-8213/ac2251}

\bibitem[{{Shu} {et~al.}(2018){Shu}, {Wang}, {Dou}, {Jiang}, {Wang}, \&
  {Wang}}]{Shu+18}
{Shu}, X.~W., {Wang}, S.~S., {Dou}, L.~M., {et~al.} 2018, \apjl, 857, L16,
  \dodoi{10.3847/2041-8213/aaba17}

\bibitem[{{Sillanpaa} {et~al.}(1988){Sillanpaa}, {Haarala}, {Valtonen},
  {Sundelius}, \& {Byrd}}]{Sillanpaa+88}
{Sillanpaa}, A., {Haarala}, S., {Valtonen}, M.~J., {Sundelius}, B., \& {Byrd},
  G.~G. 1988, \apj, 325, 628, \dodoi{10.1086/166033}

\bibitem[{{Sniegowska} {et~al.}(2022){Sniegowska}, {Grzedzielski}, {Czerny}, \&
  {Janiuk}}]{Sniegowska+22}
{Sniegowska}, M., {Grzedzielski}, M., {Czerny}, B., \& {Janiuk}, A. 2022, arXiv
  e-prints, arXiv:2204.10067, \dodoi{10.48550/arXiv.2204.10067}

\bibitem[{{Soltan}(1982)}]{Soltan82}
{Soltan}, A. 1982, \mnras, 200, 115, \dodoi{10.1093/mnras/200.1.115}

\bibitem[{{Song} {et~al.}(2020){Song}, {Shu}, {Sun}, {Xue}, {Jin}, {Zhang},
  {Jiang}, {Dou}, \& {Wang}}]{Song+20}
{Song}, J.~R., {Shu}, X.~W., {Sun}, L.~M., {et~al.} 2020, \aap, 644, L9,
  \dodoi{10.1051/0004-6361/202039410}

\bibitem[{{Stone} {et~al.}(2013){Stone}, {Sari}, \& {Loeb}}]{Stone+13}
{Stone}, N., {Sari}, R., \& {Loeb}, A. 2013, \mnras, 435, 1809,
  \dodoi{10.1093/mnras/stt1270}

\bibitem[{{Stone} {et~al.}(2018){Stone}, {Generozov}, {Vasiliev}, \&
  {Metzger}}]{Stone+18}
{Stone}, N.~C., {Generozov}, A., {Vasiliev}, E., \& {Metzger}, B.~D. 2018,
  \mnras, 480, 5060, \dodoi{10.1093/mnras/sty2045}

\bibitem[{{Stone} \& {Metzger}(2016)}]{Stone&Metzger16}
{Stone}, N.~C., \& {Metzger}, B.~D. 2016, \mnras, 455, 859,
  \dodoi{10.1093/mnras/stv2281}

\bibitem[{{Sukov{\'a}} {et~al.}(2021){Sukov{\'a}}, {Zaja{\v{c}}ek}, {Witzany},
  \& {Karas}}]{Sukova+21}
{Sukov{\'a}}, P., {Zaja{\v{c}}ek}, M., {Witzany}, V., \& {Karas}, V. 2021,
  \apj, 917, 43, \dodoi{10.3847/1538-4357/ac05c6}

\bibitem[{{Sun} {et~al.}(2013){Sun}, {Shu}, \& {Wang}}]{Sun+13}
{Sun}, L., {Shu}, X., \& {Wang}, T. 2013, \apj, 768, 167,
  \dodoi{10.1088/0004-637X/768/2/167}

\bibitem[{{Syer} {et~al.}(1991){Syer}, {Clarke}, \& {Rees}}]{Syer+91}
{Syer}, D., {Clarke}, C.~J., \& {Rees}, M.~J. 1991, \mnras, 250, 505,
  \dodoi{10.1093/mnras/250.3.505}

\bibitem[{{Tagawa} \& {Haiman}(2023)}]{Tagawa&Haiman23}
{Tagawa}, H., \& {Haiman}, Z. 2023, arXiv e-prints, arXiv:2304.03670,
  \dodoi{10.48550/arXiv.2304.03670}

\bibitem[{{Valtaoja} {et~al.}(2000){Valtaoja}, {Ter{\"a}sranta}, {Tornikoski},
  {Sillanp{\"a}{\"a}}, {Aller}, {Aller}, \& {Hughes}}]{Valtaoja+00}
{Valtaoja}, E., {Ter{\"a}sranta}, H., {Tornikoski}, M., {et~al.} 2000, \apj,
  531, 744, \dodoi{10.1086/308494}

\bibitem[{{Valtonen} {et~al.}(2012){Valtonen}, {Ciprini}, \&
  {Lehto}}]{Valtonen+12}
{Valtonen}, M.~J., {Ciprini}, S., \& {Lehto}, H.~J. 2012, \mnras, 427, 77,
  \dodoi{10.1111/j.1365-2966.2012.21861.x}

\bibitem[{{Valtonen} {et~al.}(2008){Valtonen}, {Lehto}, {Nilsson}, {Heidt},
  {Takalo}, {Sillanp{\"a}{\"a}}, {Villforth}, {Kidger}, {Poyner}, {Pursimo},
  {Zola}, {Wu}, {Zhou}, {Sadakane}, {Drozdz}, {Koziel}, {Marchev}, {Ogloza},
  {Porowski}, {Siwak}, {Stachowski}, {Winiarski}, {Hentunen}, {Nissinen},
  {Liakos}, \& {Dogru}}]{Valtonen+08}
{Valtonen}, M.~J., {Lehto}, H.~J., {Nilsson}, K., {et~al.} 2008, \nat, 452,
  851, \dodoi{10.1038/nature06896}

\bibitem[{{van Velzen}(2018)}]{vanVelzen18}
{van Velzen}, S. 2018, \apj, 852, 72, \dodoi{10.3847/1538-4357/aa998e}

\bibitem[{{van Velzen} {et~al.}(2019){van Velzen}, {Stone}, {Metzger},
  {Gezari}, {Brown}, \& {Fruchter}}]{vanVelzen+19}
{van Velzen}, S., {Stone}, N.~C., {Metzger}, B.~D., {et~al.} 2019, \apj, 878,
  82, \dodoi{10.3847/1538-4357/ab1844}

\bibitem[{{Vokrouhlicky} \& {Karas}(1993)}]{Vokrouhlicky&Karas93}
{Vokrouhlicky}, D., \& {Karas}, V. 1993, \mnras, 265, 365,
  \dodoi{10.1093/mnras/265.2.365}

\bibitem[{{{\v{S}}ubr} \& {Karas}(1999)}]{Subr&Karas99}
{{\v{S}}ubr}, L., \& {Karas}, V. 1999, \aap, 352, 452,
  \dodoi{10.48550/arXiv.astro-ph/9910401}

\bibitem[{{Wang} {et~al.}(2022){Wang}, {Yin}, {Ma}, \& {Wu}}]{Wang+22}
{Wang}, M., {Yin}, J., {Ma}, Y., \& {Wu}, Q. 2022, \apj, 933, 225,
  \dodoi{10.3847/1538-4357/ac75e6}

\bibitem[{{Weaver}(1976)}]{Weaver76}
{Weaver}, T.~A. 1976, \apjs, 32, 233, \dodoi{10.1086/190398}

\bibitem[{{Webbe} \& {Young}(2023)}]{Webbe&Young23}
{Webbe}, R., \& {Young}, A.~J. 2023, \mnras, 518, 3428,
  \dodoi{10.1093/mnras/stac3318}

\bibitem[{{Wevers} {et~al.}(2022){Wevers}, {Pasham}, {Jalan}, {Rakshit}, \&
  {Arcodia}}]{Wevers+22}
{Wevers}, T., {Pasham}, D.~R., {Jalan}, P., {Rakshit}, S., \& {Arcodia}, R.
  2022, \aap, 659, L2, \dodoi{10.1051/0004-6361/202243143}

\bibitem[{{Wevers} {et~al.}(2023){Wevers}, {Coughlin}, {Pasham}, {Guolo},
  {Sun}, {Wen}, {Jonker}, {Zabludoff}, {Malyali}, {Arcodia}, {Liu}, {Merloni},
  {Rau}, {Grotova}, {Short}, \& {Cao}}]{Wevers+23}
{Wevers}, T., {Coughlin}, E.~R., {Pasham}, D.~R., {et~al.} 2023, \apjl, 942,
  L33, \dodoi{10.3847/2041-8213/ac9f36}

\bibitem[{{Xian} {et~al.}(2021){Xian}, {Zhang}, {Dou}, {He}, \&
  {Shu}}]{Xian+21}
{Xian}, J., {Zhang}, F., {Dou}, L., {He}, J., \& {Shu}, X. 2021, \apjl, 921,
  L32, \dodoi{10.3847/2041-8213/ac31aa}

\bibitem[{{Xin} {et~al.}(2023){Xin}, {Haiman}, {Perna}, {Wang}, \&
  {Ryu}}]{Xin+23}
{Xin}, C., {Haiman}, Z., {Perna}, R., {Wang}, Y., \& {Ryu}, T. 2023, arXiv
  e-prints, arXiv:2303.12846, \dodoi{10.48550/arXiv.2303.12846}

\bibitem[{{Yao} {et~al.}(2023){Yao}, {Ravi}, {Gezari}, {van Velzen}, {Lu},
  {Schulze}, {Somalwar}, {Kulkarni}, {Hammerstein}, {Nicholl}, {Graham},
  {Perley}, {Cenko}, {Stein}, {Ricarte}, {Chadayammuri}, {Quataert}, {Bellm},
  {Bloom}, {Dekany}, {Drake}, {Groom}, {Mahabal}, {Prince}, {Riddle},
  {Rusholme}, {Sharma}, {Sollerman}, \& {Yan}}]{Yao+23}
{Yao}, Y., {Ravi}, V., {Gezari}, S., {et~al.} 2023, arXiv e-prints,
  arXiv:2303.06523.
\newblock \doarXiv{2303.06523}

\bibitem[{{Zalamea} {et~al.}(2010){Zalamea}, {Menou}, \&
  {Beloborodov}}]{Zalamea+10}
{Zalamea}, I., {Menou}, K., \& {Beloborodov}, A.~M. 2010, \mnras, 409, L25,
  \dodoi{10.1111/j.1745-3933.2010.00930.x}

\bibitem[{{Zentsova}(1983)}]{Zentsova83}
{Zentsova}, A.~S. 1983, \apss, 95, 11, \dodoi{10.1007/BF00661152}

\bibitem[{{Zhao} {et~al.}(2022){Zhao}, {Wang}, {Zou}, {Wang}, \&
  {Dai}}]{Zhao+22}
{Zhao}, Z.~Y., {Wang}, Y.~Y., {Zou}, Y.~C., {Wang}, F.~Y., \& {Dai}, Z.~G.
  2022, \aap, 661, A55, \dodoi{10.1051/0004-6361/202142519}

\end{thebibliography}


\end{document}